\documentclass[a4paper,fleqn,usenatbib,useAMS]{mnras}  

\usepackage[T1]{fontenc}
\usepackage{ae,aecompl}
\usepackage{graphicx,amsmath,amssymb}

\newcommand \hii{\ion{H}{ii}} 
\newcommand \hi{\ion{H}{i}} 
\newcommand \radm{rad~m$^{-2}$}

\title[The S-PASS/ATCA survey]{S-PASS/ATCA: a window on the magnetic universe in the southern hemisphere}

\author[Schnitzeler et al.]{
D.~H.~F.~M.~Schnitzeler$^{1,2}$\thanks{dschnitzeler@gmail.com},
E.~Carretti$^{3,4,5}$,
M.~H.~Wieringa$^3$,
B.~M.~Gaensler$^{6}$,
\newauthor
M.~Haverkorn$^{7}$,
S.~Poppi$^4$
\\ 
$^1$ Max Planck Institut f\"ur Radioastronomie, Auf dem H\"ugel 69, 53121 Bonn, Germany\\
$^2$ Bendenweg 51, 53121 Bonn, Germany\\
$^3$ Australia Telescope National Facility, CSIRO Astronomy and Space Science, PO Box 76, Epping, NSW 1710, Australia\\
$^4$ INAF - Osservatorio Astronomico di Cagliari, Via della Scienza 5, 09047 Selargius (CA), Italy\\
$^5$ INAF - Istituto di Radioastronomia, Via Gobetti 101, 40129 Bologna, Italy\\
$^{6}$ Dunlap Institute for Astronomy and Astrophysics, University of Toronto, Toronto, ON M5S 3H4, Canada\\
$^{7}$ Department of Astrophysics/IMAPP, Radboud University, PO Box 9010, 6500 GL Nijmegen, The Netherlands\\
}

\date{Accepted 2018 December 31. Received 2018 December 29; in original form 2018 July 28}

\pubyear{}

\begin{document}
\label{firstpage}
\pagerange{\pageref{firstpage}--\pageref{lastpage}}
\maketitle

\begin{abstract}
We present S-PASS/ATCA, the first wide-band radio polarimetry survey of compact sources in the southern sky. 
We describe how we selected targets for observations with the Australia Telescope Compact Array (ATCA) in the 16 cm band (1.3 -- 3.1 GHz), our observing and calibration strategy, how we analysed the data, and how we tested the quality of the data. 
The data are made publicly available.
The survey contains on average one source per five square degrees and has an angular resolution at 2.2~GHz of $\sim$ 2\arcmin$\times$1\arcmin. 
Sources with $|$RM$|$s $>$ 150~\radm\ are seen towards the Galactic plane and bright \hii\ regions, but are rare elsewhere on the sky.
Sightlines that are separated by up to 3\arcmin\ show very similar RMs. Based on this observation, we argue that the Galactic foreground is the dominant contributor to RM, confirming previous results, and that the sources must have very simple distributions of Faraday-rotating and synchrotron-emitting media.
Many sources that emit at a single RM have a spectral index in linear polarization that is (very) different from the spectral index in Stokes $I$. Analysing ratios of flux densities $Q/I$ and $U/I$ (to correct for spectral index effects) then leads to erroneous results.
About 80 per cent of sightlines in our survey are dominated by emission at only one RM. Therefore, RMs that were determined previously from narrow-band observations at these frequencies are still safe to use.
\end{abstract}

%

\section{Introduction}\label{introduction.sec}
Understanding the role that magnetic fields play in galaxies, and how these magnetic fields formed in the early universe and developed over time, are perhaps the two most important questions in magnetic field research. Several techniques are employed to answer these questions. We focus on the Faraday effect, which is used to study the properties of magnetic fields in ionized gas. The magnitude of the Faraday effect is described by $\chi - \chi_0$ = RM$\lambda^2$, with
\begin{eqnarray}
\mathrm{RM}\, \left(\mathrm{rad~m}^{-2}\right)\ \approx\ 0.81 \int_\mathrm{source}^\mathrm{observer} n_\mathrm{e} B_\| \mathrm{d}l\, .
\label{rm_definition.eqn}
\end{eqnarray}
Here $\chi$ and $\chi_0$ are the observed and emitted position angles, respectively, of a linearly polarized radio wave (in radians), $\lambda$ the observing wavelength (m), $n_\mathrm{e}$ the free electron density (cm$^{-3}$), $B_\|$ the length of the magnetic field vector projected along the line of sight ($\mu$G), and d$l$ is an infinitesimal distance interval along the line of sight (pc).
The rotation measure (RM) encapsulates the physical properties of the intervening medium, and it contains contributions from the Earth's ionosphere, our own Milky Way, distant galaxies and galaxy clusters, and the cosmic web. 

The Milky Way is a prime laboratory for studying the physical effects of magnetic fields over a wide range of physical scales that are not accessible in external galaxies (see, e.g., \citealt{simonetti1984}, \citealt{minter1996}). Before the year 2000, RMs had been determined for about a thousand sources by, e.g., \cite{gardner1969}, \cite{vallee1975}, and \cite{simard-normandin1981}, who derived RMs based on measurements at only a few frequencies. 
Such measurements have led to the insight that our own Milky Way could well be the dominant contributor to the RMs that are measured for extragalactic radio sources (see Section~\ref{intrinsic_properties.sec}). This makes it vital to subtract the Galactic foreground in studies of the intrinsic properties of extragalactic radio sources. The RM in the rest frame of the source is larger than the RM we observe by a factor of $(1+z)^2$, where $z$ is the redshift of the source. Therefore, when calculating rest-frame RMs for sources at high redshift, any residual Galactic foreground RM is amplified by the factor $(1+z)^2$ to a large number. 

The RM catalogue of \cite{taylor2009} (`TSS09'), which includes 37,543 RMs that the authors derived from polarization measurements in the NRAO VLA Sky Survey (NVSS, \citealt{condon1998}), provides a much finer grid of RM measurements on the sky than was previously available, of about 1 source per square degree. This enabled detailed studies of objects in our Milky Way, a more reliable estimation of the Galactic contribution of RMs measured for extragalactic sources, and investigations of RMs associated with distant radio sources (see Section~\ref{science.sec}).
However, the RMs in the catalogue by Taylor et al. were derived based on measurements at only two frequencies, and do not cover the southern sky below a declination of -40\degr. 
Since the turn of the century, wide-band receivers have been installed on radio telescopes, which made it possible to measure all Stokes parameters for a wide range of frequencies simultaneously. This has led to new insights on the physical properties of radio sources and their magnetic fields, and a better understanding of how Faraday-rotating and synchrotron-emitting media are distributed in radio sources, including our own Milky Way (see, e.g., \citealt{osullivan2012}, \citealt{schnitzeler2009}, \citealt{vaneck2017}).

We present S-PASS/ATCA, the first wide-band polarimetric survey of radio sources over a large region on the sky. 
Each target field in S-PASS/ATCA contains data for Stokes $I$, $Q$, and $U$ between $\approx$ 1.3--3.1 GHz, sampled with 8~MHz channels, a major improvement over existing data. 
Furthermore, S-PASS/ATCA sightlines are spread over the entire sky south of declination = 0\degr , filling in the gap in the southern sky that was not covered by RMs from the catalogue by Taylor et al. 
S-PASS/ATCA builds on S-PASS, the S-band Polarization All-Sky Survey, which mapped the sky below a declination of 0\degr\ with the Parkes radio telescope at a frequency of 2.3 GHz (\citealt{carretti2010, carretti2013, carretti2019}). However, the S-PASS survey has a narrow bandwidth of 256 MHz, which, at these frequencies, makes it less than ideal for measuring Faraday rotation.
We observed suitable candidates that we selected from S-PASS with the Australia Telescope Compact Array (ATCA), a six-element radio interferometer close to Narrabri, New South Wales, Australia.
After the Compact Array Broadband Backend (CABB) upgrade \citep{wilson2011}, the ATCA nowadays observes routinely in the 1.3 -- 3.1 GHz (`16 cm') band; this frequency band has been proven to be excellent for studying Faraday rotation (see, e.g., \citealt{osullivan2012}).
With the increased bandwidth provided by the 16 cm band we can not only determine RMs more accurately than was possible with the original Parkes data, we can also identify sources that emit at more than one RM.
The observations for S-PASS/ATCA attain a resolution in RM, RM$_\mathrm{Rayleigh}$ = 71~\radm\ (using equation~14 in \citealt[`S18']{schnitzeler2018}, and assuming a contiguous frequency coverage; in practice, radio-frequency interference makes frequencies between 1.5 -- 1.6 GHz unusable, see also Fig.~\ref{example_los.fig}). The full width at half-maximum of the RM spread function is about 1.2 times larger than RM$_\mathrm{Rayleigh}$, or 86~\radm\ for our observations.
Re-observing candidates from the Parkes survey with the ATCA also improves the angular resolution, 
from 9\arcmin\ (FWHM) to about 2\arcmin $\times$1\arcmin\ (Section~\ref{post-cal.sec}).
To keep the data volume low, we increased the width of the frequency channels from 1 MHz (their native resolution) to 8 MHz. This reduces our ability to detect sources with very large (positive or negative) RMs. Based on the analysis in \cite{schnitzeler2015}, we estimate that sources with $|$RM$|$ $\approx$ 11,400~\radm\ are detected with only half the flux density they emit.

This paper is structured as follows. We explain in Section~\ref{observations.sec} how we selected candidates for observations with the ATCA, and we describe our observing strategy. In Section~\ref{calibration-post-processing.sec} we outline how we calibrated these data, and how we extract information on the polarization properties of the radio sources. We test how well our new results compare to results that have been published in the literature previously. In Section~\ref{quality.sec} we describe these tests. We analyse what these new results tell us about RMs produced in the Milky Way and in the sources themselves in Section~\ref{science.sec}. In Section~\ref{access.sec} we list which data products are made available online, and where these can be accessed. We summarise our analysis in Section~\ref{conclusions.sec}.

Throughout this paper, we will write the linear polarization vector as $\bmath{L} = Q + \mathrm{i}U$, and we will use the nomenclature introduced in Appendix~A of \cite{schnitzeler2017}. The sign of the flux density spectral index, $\alpha$, is defined such that the spectrum of a source can be written as $S_\nu = S_0\left(\nu/\nu_\mathrm{ref}\right)^\alpha$, where $\nu$ is the observing frequency.

\section{Observations}\label{observations.sec}

\subsection{Selection of targets in the Parkes survey}\label{parkes_candidate_selection.sec}
We applied a median filter to S-PASS images of Stokes~$I$, $Q$, and $U$ to filter out diffuse emission on angular scales $\gtrsim$ 20\arcmin\ (see also \citealt{lamee2016}). Then we identified point sources using the \textsc{miriad} task \textsc{sfind} (\citealt{sault1995}, \citealt{hopkins2002}), extracted Stokes~$Q$ and $U$ frequency spectra for each point source, and used RM synthesis (\citealt{burn1966}, \citealt{brentjens2005}) to identify polarized emission at RMs between $\pm$~1000~\radm .
We did not correct for spectral index effects or a variation in the sensitivity across the frequency band when we ran RM synthesis.
We identified suitable candidates for follow-up observations with the ATCA based on the following criteria:
\begin{enumerate}
\item polarized flux density $>$ 5 mJy at 2.3 GHz, which corresponds to a signal-to-noise ratio of at least five at 2.3 GHz,
\item a polarization fraction $>$ 1 per cent, to avoid instrumental polarization artefacts (\citealt{carretti2019}), 
\item each source must have a total intensity counterpart in the NRAO VLA Sky Survey \citep[NVSS, ][]{condon1998}, the Sydney University Molonglo Sky Survey (SUMSS, \citealt{bock1999}, \citealt{mauch2003}), or the Molonglo Galactic Plane Survey (MGPS, \citealt{green1999}, \citealt{murphy2007}).
\end{enumerate}
The angular resolution of the NVSS, SUMSS, and MGPS (all $\sim$~45\arcsec) is much higher than the angular resolution of the S-PASS data collected with the Parkes telescope, allowing a reliable identification of counterparts.

This way, we selected 5102 candidates for follow-up observations with the ATCA.
None of these targets lie within $\approx$1.5\degr\ of the Galactic plane. 
However, RMs in the Galactic plane region are provided by the Southern Galactic Plane Survey (SGPS, \citealt{haverkorn2006}) and by \cite{vaneck2011}.

\begin{figure}
\begin{centering}
\resizebox{\hsize}{!}{\includegraphics[]{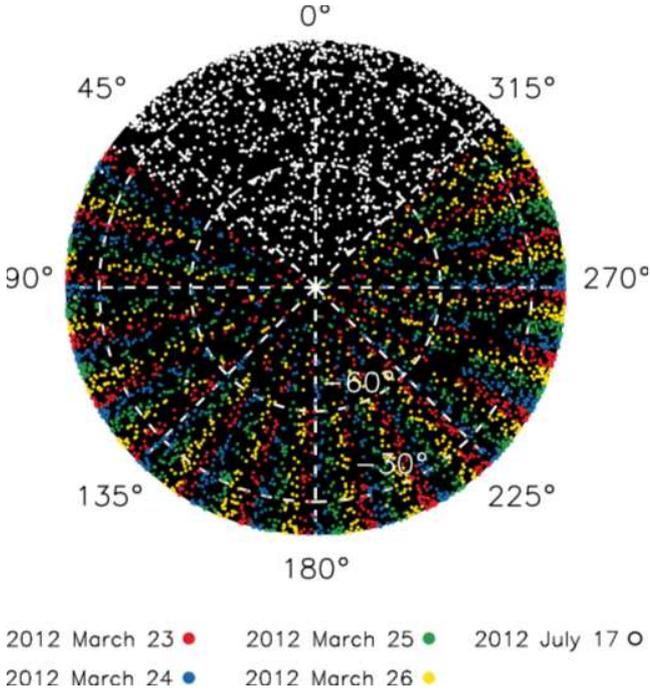}}
\caption{
Target selection for each of the five observing dates in 2012. Each dot represents a single candidate from the S-PASS single-dish survey. The coordinate system is centred on the south equatorial pole. Right Ascension is expressed in units of degrees.
}
\label{strips.fig}
\end{centering}
\end{figure}

\subsection{Pilot project}\label{observations_pilot_project.sec}
On the 10th of March 2011, we observed 118 S-PASS targets with the ATCA in the 1.5A configuration (east-west array, baselines between 153 - 4469 m), as a pilot study for the larger S-PASS/ATCA survey. 
We used this pilot to test if we can observe a large number of sources in a short period of time, and to inspect the quality of the data. For this purpose, the ATCA did not have to be in the same configuration as the one we used for our survey.
The targets lie between 231\degr $<$ $l$ $<$ 246\degr\ and -50\degr $<$ $b$ $<$ 0\degr, which places them in-between the bright \hii\ emission regions of the Gum nebula and the Orion molecular cloud complex (see Figs.~\ref{rms_pilot.fig} - \ref{finder_chart.fig}).
We observed PKS B1934-638 at the start of this run for 5 minutes. The secondary calibrator, PKS B0614-349, and our target fields were observed multiple times during the night, typically five-six times. Most sources were observed for about 90 -- 120 seconds in total.

\subsection{Survey}\label{atca_observing_strategy.sec}
We used the ATCA in the hybrid array configuration H168 to observe sources from the S-PASS/ATCA survey (baselines between 61 - 192 m, excluding baselines to antenna 6).
The hybrid configuration of the ATCA has antennas not only on an east-west track but also on a short north-south track. 
Using a hybrid configuration improves the {\it uv}-coverage of the observations, which is important in particular for those candidates that lie close to the celestial equator.
We targeted 4563 candidates in 77 hours of observing time. 
Each candidate was observed for 36 seconds, slewing between targets took 12 seconds typically.
We observed mostly at night, to minimize the impact of radio-frequency interference and variations in the ionospheric RM.
To avoid observing targets within $\lesssim$ 52\degr\ from the Sun, targets were divided into two groups that were observed between 2012 March 23 to 26 and on July 17.
The surface density of candidates selected from the Parkes survey increases closer to the Galactic plane. This is noticeable particularly at Galactic latitudes between $\pm$~40\degr.
To make the surface density more uniform, and to save observing time, we divided this region into bins that measure 5\degr $\times$ 5\degr\ on the sky, and selected at random 8-9 candidates in each  bin for follow-up observations with the ATCA.

To observe so many targets over such a large area, we divided the sky into narrow strips in Right Ascension that have a width of 15 minutes in RA (Fig.~\ref{strips.fig}).
Because the surface densities of candidates were slightly different on days three and four, strips on day three are a bit narrower than 15 minutes in RA, and on day four a bit wider, so that the total observing time for each strip is about the same.
Slew times between targets in each strip are minimized by solving the travelling salesman problem, for which we used the $\textsc{miriad}$ program $\textsc{atmos}$.
In runs 1-4, observations started close to Declination 0\degr, then moved in the direction of the south equatorial pole.
Observations of the next strip started close to Dec. -~90\degr, and continued towards the celestial equator. This way, during each observing run the array slews up and down in declination to observe targets in different strips.
In run 5, observations started close to the south equatorial pole and then moved in the direction of the celestial equator.
Completing the observations of a single strip typically takes a bit over one hour, during which the sky rotates towards the west.
The RA of the next strip is one hour higher than the RA of the previous strip, so that the array slews back towards the east when it starts observing a new strip.
Because the observing time for each strip is about equal to the change in RA between strips, the observations are almost able to keep up with the changing position of each strip on the sky. 
This means that the array will progress from observing strips in the east (at the start of each observing run) to the west (at the end of each run) only slowly, allowing us to cover a wide range in RA each run.
We did not re-visit targets, with some exceptions.

\begin{table*}
\centering
\caption{
Overview of sources that we used to calibrate gain amplitudes and phases. For each source we list an alias, whether the source was listed as a Gigahertz Peaked Spectrum (GPS) or Compact Steep Spectrum (CSS) source by \citet{edwards2004} or \citet{randall2011}, and its equatorial coordinates, taken from the ATCA calibrator data base http://www.narrabri.atnf.csiro.au/calibrators/. 
We also present its S-PASS/ATCA flux densities at the reference frequency (2.1 GHz) in Stokes~$I$ and in $L$, its RM, and during which runs the source was observed.
Runs 1-4 occurred between 2012 March 23-26, run 5 on 2012 July 17.
$L$ and RM are listed for the brightest source component that we fitted to Stokes~$Q$ and $U$, we used the subscript `1' to indicate this. 
If calibrators have been observed on more than one day, we list weighted means for $I$, $L$, and RM, using the uncertainties in the measurements as weights. Measurements for individual runs are presented in Appendix~\ref{Appendix_B.sec}.
RMs were not corrected for ionospheric Faraday rotation.
}
\label{calibrators.tab}
\begin{tabular}{lcccccccccccc}
\hline
Name			& Alias		& Type 		& RA (J2000)	& Dec. (J2000)	& $I_\mathrm{ref}$ 	& $L_\mathrm{1,ref}$	& RM$_1$	& \multicolumn{5}{c}{observing run}\\
				&			&			& hh:mm:ss	& dd:am:as 	& mJy		& mJy					& \radm		& 1		& 2		& 3		& 4		& 5	\\  
\hline
PKS B0008-421	&			&			& 00:10:52	& -41:53:11	& 3240		& 10			& 2			&		&		&		&		& $\star$	\\
PKS B0023-263        & OB-238		& CSS		& 00:25:49	& -26:02:13	& 6728		& 25			& 4			&		&		&		&		& $\star$	\\
PKS B0237-233 	&			& GPS		& 02:40:08	& -23:09:16	& 5080		& 116		& -3			&		&		&		&		& $\star$	\\
PKS B0403-132	& OF-105		& CSS		& 04:05:34	& -13:08:14	& 3339		& 70			& 13			&		&		&		&		& $\star$	\\
PKS B0420-014	&			&			& 04:23:16	& -01:20:33	& 3049		& 59			& -44		& $\star$ 	& $\star$	& $\star$	& $\star$	& 		\\
PKS B0440-003 	&			&			& 04:42:39	& -00:17:43	& 3579		& 70			& 57			& $\star$ 	&		&		&		&		\\
PKS B0537-441	&			&			& 05:38:50	& -44:05:09	& 7473		& 225		& 60			&		& $\star$	&		& $\star$	&		\\
PKS B0607-157	&			&			& 06:09:41	& -15:42:41	& 3048		& 97			& 69			&		&		& $\star$	& $\star$	&		\\
PKS B0823-500 	&			&			& 08:25:27	& -50:10:38	& 6031		& 20			& 231 (360$^1$) & $\star$ & $\star$	& $\star$	& $\star$	& $\star$	\\
PKS B1127-145	&			&			& 11:30:07	& -14:49:27	& 4280		& 159		& 42			& $\star$  & $\star$	& $\star$	& $\star$	&		\\
PKS B1215-457 	&			& CSS		& 12:18:06	& -46:00:29	& 3846		& 17			& -117		& $\star$  & $\star$	& 		&		&		 \\

PKS B1308-220$^2$ & 3C283		& CSS		& 13:11:39	& -22:16:42	& 3329		& 3 (111$^2$)	& 1			& $\star$  & $\star$	& $\star$	& $\star$	&		\\

PKS B1421-490	&			& CSS		& 14:24:32	& -49:13:50	& 7306		& 81$^3$		& 32$^3$		& $\star$  &		& $\star$	& $\star$	&		\\
PKS B1613-586 	&			&			& 16:17:18	& -58:48:08	& 4620		& 50			& 61			& $\star$  & $\star$	& $\star$	& $\star$	&		\\
PKS B1730-130	&			&			& 17:33:03	& -13:04:50	& 4570		& 221		& -60		&		&		&		& $\star$	&		\\
PKS B1827-360$^4$ &			& GPS		& 18:30:59	& -36:02:30	& 4279		& 3			& -326		& $\star$  & $\star$	& $\star$	& $\star^5$&		\\
PKS B1934-638$^4$ &			& GPS		& 19:39:25	& -63:42:46	& 12346		& 5			&  399		& $\star$  & $\star$	& $\star$	& $\star^5$& $\star$ \\
PKS B2032-350 	&			&			& 20:35:48	& -34:54:09	& 3982		& 258		& 2			&		&		& $\star$	& $\star$	& $\star$ 	\\
PKS B2203-188 	& OY-106		& CSS		& 22:06:10	& -18:35:39 	& 5468		& 39			& 5			&		&		&		&		 & $\star$ 	\\
PKS B2223-052	& 3C446		&			& 22:25:47	& -04:57:01	& 7369		& 277		& -31		&		&		&		&		 & $\star$ 	\\
\hline
\end{tabular}
\begin{list}{}{}
\item[$^1$] during run 5 the brightest and second brightest source components of PKS B0823-500 had RMs of -2~\radm\ and +353~\radm , respectively (Table~\ref{mult_days_RM1.tab}). 
The weighted mean RM of the other four runs is 360~\radm .
\item[$^2$] PKS B1308-220 was detected with a polarization fraction $L_{1,\mathrm{ref}}/I_\mathrm{ref} > 0.1$ per cent only during run 2; on the other three runs the polarization percentage of the signal was so low that it could be produced by the telescope itself. On run 2 we measured a polarized flux density of 111 mJy, the weighted mean of $L_{1,\mathrm{ref}}$ is calculated from the other three observing runs (Table~\ref{mult_days_L1ref.tab}).
\item[$^3$] the best fit for run 3 had $L > I$ and was therefore discarded.
\item[$^4$] the average polarization percentage of this source is $<$ 0.1 per cent, the measured polarized signal could therefore be purely instrumental.
\item[$^5$] in run 4, all fitted models converged on $\alpha$ = -6 or +3, and were therefore discarded
\end{list}
\end{table*}

\section{Calibration and post-processing}\label{calibration-post-processing.sec}
\subsection{Calibrating the pilot project}\label{calibration_pilot.sec}
We used PKS B1934-638 for the initial calibration of antenna delays, gains, and phases at the start of the observations, and also to calibrate the bandpass, flux density scale \citep{reynolds1994}, and polarization leakages, using the $\textsc{miriad}$ tasks $\textsc{mfcal}$ and $\textsc{gpcal}$.
PKS B1934-638 is a bright, unpolarized point source; furthermore, a deep image of the field surrounding PKS B1934-638 that was created by E.~Lenc from archival ATCA data shows no sources brighter than $\approx$ 20 mJy.
Therefore, a single short observation of this source is sufficient to calibrate antenna leakages.
We investigated the stability of the polarization leakage solution using two observations of PKS B1934-638 that are separated by 24 hours. 
We derived the calibration parameters from five minutes worth of data in the first observation, then we copied these to the second observation of this source, and measured the polarized flux density of the highest peak in the RM spectrum.
The polarization fraction of this peak is $L/I$ = 10~mJy~/~12360~mJy, or less than 0.1 per cent.
All candidates that we selected from the Parkes survey have a higher polarization fraction than that, therefore the leakage calibration is more than sufficient for our purpose.
We transferred these calibration solutions to our secondary calibrator PKS B0614-349, (which has a polarization percentage of 0.08\% at 2.1~GHz), and applied standard $\textsc{miriad}$ procedures to calibrate the complex gains using this source.
Then we transferred all calibration tables to each of the target fields, and used the S-PASS/ATCA pipeline to self-calibrate and flag each target field (we describe this procedure in the next section).

\subsection{Calibrating the survey}\label{calibration.sec}
At the start of each observing run we observed PKS B0823-500 (runs 1--4) or PKS B1934-638 (run 5) for the initial calibration of antenna delays, amplitudes, and phases.
We observed PKS B1934-638 for several minutes during each run to calibrate the bandpass, flux density scale, and polarization leakages, identical to what we described in Section~\ref{calibration_pilot.sec}.
ATCA data are normally calibrated by observing a secondary calibrator interleaved with observations of the target field(s); however, for S-PASS/ATCA this is not possible because we do not re-observe most targets. 
Furthermore, we cannot use the $\textsc{miriad}$ task $\textsc{gpcal}$, since it requires that a calibrator is observed with good parallactic angle coverage, to separate instrumental from source-intrinsic polarization effects.
Instead, first we self-calibrate all calibrators, then we self-calibrate all other targets.
Appendix~\ref{Appendix_A.sec} shows the details of this process.
Table~\ref{calibrators.tab} lists all sources that we used as calibrators. These sources were selected because they are unresolved on the baselines used in our observations, and bright, so that they dominate the field of view and source confusion is mitigated.
Throughout the self-calibration process, data were flagged automatically using the $\textsc{miriad}$ routines $\textsc{uvflag}$ and $\textsc{tvclip}$.
Shadowed baselines were flagged automatically at the start of the calibration process.

\subsection{Post-calibration, Extracting polarization information}\label{post-cal.sec}
\begin{figure}
\begin{centering}
\resizebox{0.9\hsize}{!}{\includegraphics[]{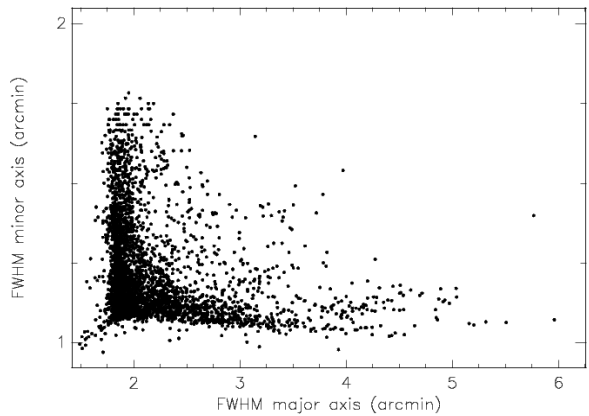}}
\caption{
Size of the restoring Gaussian beam used to create the final CLEANed image for each ATCA field, at 2.2 GHz. Note the difference in scale between the horizontal and vertical axes.
}
\label{beamsizes.fig}
\end{centering}
\end{figure}

Since our observations cover a wide range in frequency, the size of the synthesized beam changes by a large amount across the band. 
When analysing sources from the survey, we found that convolving all channels to the same beam size corrupts the shape of Stokes~$I$ frequency spectra, making them unusable (we did not test how Stokes~$Q$ and $U$ are affected). This might be related to the poor {\it uv}-coverage of our observations. Therefore we did not modify the data to make the beams more uniform. Given the large size of the synthesized beam, $\sim 2\arcmin\times1\arcmin$, see Fig.~(\ref{beamsizes.fig}), most sources are unresolved, in which case the change in size of the synthesized beam will not affect our results.

We used the $\textsc{miriad}$ task $\textsc{sfind}$ to create a list of sources in a Stokes $I$ image of each target field, and we identified sources by cross-correlating the brightest two sources in Stokes $I$ in each field with the PKSCAT90\footnote{http://vizier.u-strasbg.fr/viz-bin/VizieR?-source=VIII/15} \citep{wright1990} and NVSS catalogues (if a match was found with a source from PKSCAT90, it was not also correlated with NVSS). Some target fields contain more than two sources in Stokes $I$; in those cases we limited our analysis to the brightest two sources.
If no counterpart was found, we named a source after its sky coordinates: SPASS Jhhmmss$\pm$ddmmss. Seconds and arcseconds are truncated, not abbreviated, following the naming convention of NVSS sources.

For the two sources that are brightest in Stokes $I$ in each field we extract Stokes $I$, $Q$, and $U$ flux densities as a function of frequency, together with the uncertainties in these measurements. 
We extract flux densities directly from the {\it uv}-visibilities, using the $\textsc{miriad}$ task $\textsc{uvspec}$ (extracting only the real parts of the visibilities), instead of creating CLEANed channel maps. This saves time and therefore greatly increases processing speed, and no additional storage space is needed when saving the channel maps.
In the case of sources from the pilot project, we include only baselines with a projected length larger than 1 k$\lambda$.

We analyse the data using the $\textsc{Firestarter}$ program, a QU-fitting based algorithm that we describe in S18.
This program can be downloaded from the following URL\footnote{https://github.com/dschnitzeler/firestarter}.
$\textsc{Firestarter}$ fits for the spectral index of each source component (assuming the source emits a power law synchrotron spectrum, before this emission is depolarized), and includes all available information on the measurement uncertainties. In particular, it does not assume that the noise variances in Stokes $Q$ and $U$ are equal and constant across the band.
As we discussed in \cite{schnitzeler2017} and S18, this makes the program much more capable at handling real observations and real sources than competing algorithms like RM~synthesis.
The user specifies a list of model components that $\textsc{Firestarter}$ should fit to the data, together with the maximum number of allowed model components.
Complex models are fitted iteratively, starting with the simplest possible model, and subsequently adding new components.
We fit data from the pilot project and from the survey with up to five point sources in RM, without implementing a cut-off in the reduced $\chi^2$. Each point source is modelled using equation~3 in S18.
At the heart of $\textsc{Firestarter}$ lies the Levenberg-Marquardt optimalisation algorithm (\citealt{levenberg1944}, \citealt{marquardt1963}, \citealt{more1978}, \citealt{markwardt2009}), which, in the case of S-PASS/ATCA, needs to be initialized with a starting value for the spectral index $\alpha$ of the source and its RM. As starting values we use $\alpha$ = 0 and the RM that shows the highest polarized flux density in an RM spectrum calculated between $\pm$ 2500~\radm.
$\textsc{Firestarter}$ ranks models that were fitted to the data automatically, and calculates the detection significance and signal-to-noise ratio, relying on concepts from statistics and information theory. 
We selected the Bayesian Information Criterion, without applying model averaging, for ranking model fits.
In S18 we showed that this is the most selective criterion for observations that cover the frequency range of S-PASS/ATCA.
We discard fits from the final catalogue in three cases: (1) if a fit has converged on a spectral index that is -6 or +3, the extreme values allowed by the fitting procedure, (2) if a fit is flagged during the fitting process, or (3) if the brightest polarized source ($L_{1,\mathrm{ref}}$) has a polarization percentage larger than 100 per cent. Because $\textsc{Firestarter}$ does not fit models to Stokes $I$, the latter check is carried out after $\textsc{Firestarter}$ has been applied.
For sources that were observed on more than one day, we list the observation with the highest polarized signal-to-noise ratio for $L_{1,\mathrm{ref}}$.

\begin{figure}
\begin{centering}
\resizebox{\hsize}{!}{\includegraphics{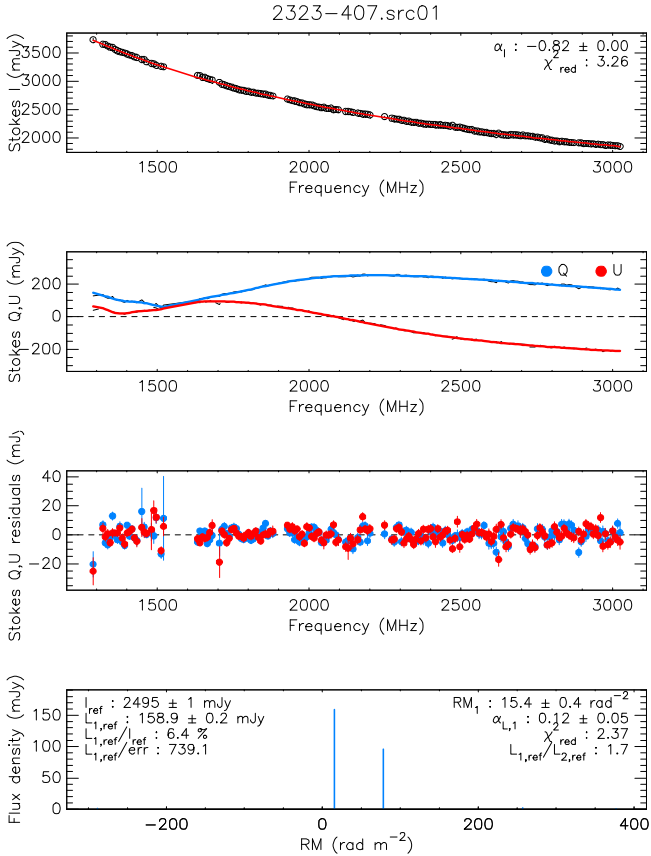}}
\caption{
S-PASS/ATCA data together with the best-fitting models for the radio source PKS~B2323-407. 
Diagnostic plots like this one are available for the two sources that are brightest in Stokes~$I$ in each target field.
The top three panels show frequency spectra for Stokes~$I$ (top), $Q$ and $U$ (second panel, showing data plus best-fitting model), and the residuals in Stokes~$Q$ and $U$ after subtracting the best-fitting model (third panel).
The numbers in the top panel show the spectral index of the power-law fitted to the Stokes~$I$ data and the reduced chi-squared of this fit. The error bars indicate 1$\sigma$ errors on the measurements.
The panel at the bottom shows the sources that were fitted to the $Q,U$ data.
In this panel, numbers in the column on the left are the fitted Stokes~$I$ flux density at the reference frequency (2.1 GHz), the polarized flux density of the brightest source component, $L_{1,\mathrm{ref}}$, 
the polarization percentage, and the polarized signal-to-noise ratio (calculated using the method outlined in Appendix~A of S18). 
In the column on the right, RM$_1$ and $\alpha_\mathrm{L,1}$ are the RM and polarized flux density spectral index of the brightest source component, $\chi^2_\mathrm{red}$ the reduced chi-squared of the model fit to the Stokes~$Q$ and $U$ frequency spectra, and $L_{1,\mathrm{ref}}/L_{2,\mathrm{ref}}$ the ratio between the polarized flux densities of the brightest and second brightest source components.
}
\label{example_los.fig}
\end{centering}
\end{figure}

As an illustration, we show in Fig.~(\ref{example_los.fig}) the calibrated data and model fits for the radio source PKS~B2323-407. The program identifies bright polarized emission at two RMs, and faint polarized emission of about 2~mJy at three additional RMs (bottom panel). The fit residuals to the Stokes $Q$, $U$ data have a reduced $\chi^2$ of 2.4. The Stokes~$I$ spectrum can be fitted by a single power law (with a reduced $\chi^2$ of 3.2), and it is clear that the spectral index fitted to Stokes~$I$, -0.82, is very different from the spectral index fitted to the brightest source component in polarized flux density, where $\alpha_{\mathrm{L},1}=0.12$. 
We will investigate this for a larger sample of sources in Section~\ref{intrinsic_properties.sec}.

\begin{figure}
\begin{centering}
\resizebox{0.85\hsize}{!}{\includegraphics[trim={0cm 0cm 0 0}, clip]{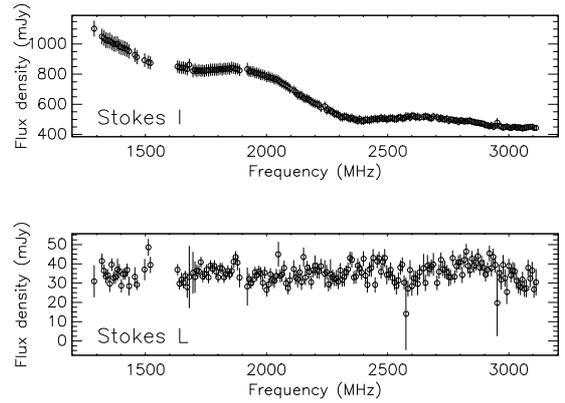}}
\caption{
Stokes $I$ and $L$ frequency spectra for the brightest source in Stokes $I$ in the target field of PKS B1923-328 (top and bottom panel, respectively). The bump in the Stokes $I$ spectrum is due to confusion with another source in the field of view, but this bump is missing from the polarization data. Error bars indicate 1$\sigma$ uncertainties. We used standard error propagation to calculate the errors in $L$, allowing for the measurement uncertainties in Stokes $Q$ and $U$ to be different in each frequency channel.
}
\label{confusion.fig}
\end{centering}
\end{figure}

Finally, we want to mention two issues that we identified in the S-PASS/ATCA data.
First, if sources are bright, the models fitted to the polarization data can show large values for the reduced $\chi^2$, indicating that we are not fitting the correct models to the data, that the deviations between the fitted models and the data are not described by Gaussian noise (for example, they are produced by calibration errors), or a combination of the two. 
In fainter sources, these effects can be hidden in the noise. 
If the second explanation is responsible for the large values of the reduced $\chi^2$, then equation~(2) from S18 does not describe the likelihood. Quantities that depend on the likelihood, like the detection significance and the Bayesian Information Criterion, then are also not reliable.
However, in Section~\ref{quality.sec} we show that our results match those from previous observations in the (vast) majority of cases that we tested.
Future projects can clarify the origin of the large $\chi^2$ values, for example, by observing at a higher angular resolution.

Secondly, our S-PASS/ATCA observations could be affected by source confusion due to aliasing in the dirty beam, since we extract Stokes~$I$, $Q$, and $U$ frequency spectra directly from the visibilities, instead of making CLEAN-ed channel maps.
Source confusion can manifest itself as oscillations on top of a power-law spectrum in Stokes~$I$, but not necessarily in Stokes $Q$ and $U$, as Fig.~(\ref{confusion.fig}) shows (e.g., because the confusing source may be unpolarized). 
As a result of this, the power law that we fit to the Stokes~$I$ spectrum can have a large value for the reduced $\chi^2$, and polarized emission can occur at more than one RM.
However, a large value for the reduced $\chi^2$ should not be used to identify automatically which fields are affected by source confusion.
There can be other reasons why this value is high: for example, the Stokes~$I$ spectrum of PKS B0823-500 peaks at around 1.5-1.6~GHz, therefore this spectrum can not be fitted well by a power law.

\section{Data quality}\label{quality.sec}
To assess the quality of the S-PASS/ATCA data, we compared Stokes $I$ and $L$ at the reference frequency (2.1~GHz), RM, and $\alpha_\mathrm{L}$ of sources that have been observed more than once.
Also, we compared RMs from the survey with RMs from the pilot project and RMs that have been published by TSS09 and by \cite{mao2010}.
In our analysis, we include only those sightlines that satisfy the following criteria: 
\begin{enumerate}
\item fitted Stokes $I$ at the reference frequency $>$ 0 mJy, 
\item polarization fraction $L/I$ between 0.1 and 100 per cent (this flux density ratio is calculated at the reference frequency),  
\item the spectral index that is fitted to Stokes $I$ must be between -6 and 3, and
\item {the brightest component that is fitted to the polarization data should be detected at least at ten times the noise level.} 
\end{enumerate}
Additionally, a visual inspection of diagnostic plots like Fig.~(\ref{example_los.fig}) showed that in many of the target fields, the second brightest source in Stokes~$I$ is not real.
Given the {\it uv}-coverage of the data, such a spurious source could be a local maximum in the synthesized dirty beam that is mistaken by CLEAN for a genuine source. Since the pattern of local maxima and minima changes considerably across the 16 cm band of the ATCA, the resulting Stokes~$I$ spectrum will show features that indicate the source is unlikely to be real.
Therefore, we include in our analysis only those sources that could be real based on the shape of their Stokes $I$ spectrum, taking into account that bumps in the Stokes~$I$ spectrum could come from confusion (see Fig.~\ref{confusion.fig}), but nevertheless belong to a real source.

\begin{figure}
\begin{centering}
\resizebox{0.9\hsize}{!}{\includegraphics[]{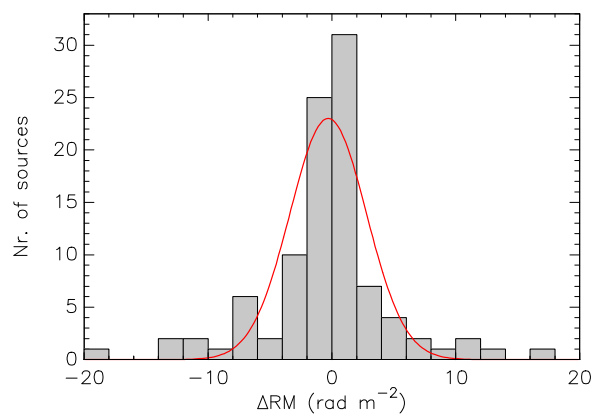}}
\caption{
Distribution of $\Delta$RM values for sources that have been observed more than once. Thirteen data points fall outside the plot range, the largest $\Delta$RM values are -360~\radm\ and 292~\radm. The red line shows a Gaussian with the mean and standard deviation reported for RM in the final two columns of Table~\ref{parameter_comparison_multiple_obs.tab}. The height of this Gaussian was chosen such that the area under the curve is the same as the number of sources used to calculate these metrics.
}
\label{Delta_RM_sources_obs_multiple_times.fig}
\end{centering}
\end{figure}

For sources that have been observed more than once, we calculate differences between the four parameters that we mentioned in the first paragraph of this section.
Also, we calculate normalized differences between these parameters. For example, the normalized difference in RM, $\Delta\mathrm{RM}\left(\mathrm{normalized}\right) = \left(\mathrm{RM}_2 - \mathrm{RM}_1\right)/\sqrt{\mathrm{err}_{\mathrm{RM},1}^2 + \mathrm{err}_{\mathrm{RM},2}^2 }$, where RM$_1$ refers to the first measurement of RM, and $\mathrm{err}_{\mathrm{RM},1}$ is the measurement uncertainty in this RM (the subscript `2' refers to the second measurement). 
If the differences between $\mathrm{RM}_1$ and $\mathrm{RM}_2$ are purely due to Gaussian noise, then the normalized $\Delta\mathrm{RM}$ follow a Gaussian distribution with a mean of zero and a standard deviation of one.
Then we determine the population means and standard deviations of the differences and normalized differences.
We select two populations of sources: only those sources that emit at one RM, and all sources. 
The results of this analysis are shown in Table~\ref{parameter_comparison_multiple_obs.tab}, and in Fig.~(\ref{Delta_RM_sources_obs_multiple_times.fig}).
The metrics show that the source parameters that we analysed can be determined accurately. However, we caution that the populations also contain outliers that have been removed by the algorithm that calculates robust statistics. For example, two observations of PKS B1442-421 show a difference in RM of 292~\radm. In this case, the large difference in RM originates because this source emits at two RMs. In the first observation one peak is the brightest, while in the second observation the other peak is brightest.
Roughly 80 per cent of the sightlines are selected to calculate the final two columns in Table~\ref{parameter_comparison_multiple_obs.tab}.
The standard deviations of normalized differences are larger than one (except for $\Delta$RM of sources that emit at one RM): this indicates that the differences in the parameters that we analysed are due not only to the measurement uncertainties.
If we consider only the parameters that we derived for sources from Table~\ref{calibrators.tab} (see Appendix~\ref{Appendix_B.sec}), and calculate standard deviations from multiple observations of each source, then we find that the standard deviations in Stokes $I$ are smaller than 60~mJy, which corresponds to $\lesssim$ 1 per cent of the mean. The exception is PKS B0823-500, which has a standard deviation in Stokes $I$ of 196 mJy (3 per cent of the mean). 
In $L$, the standard deviations can be up to 34~mJy, or 35 per cent of the mean (not counting the detection of PKS B1308-220 on run 2).

\begin{table}
\centering
\caption{
The population mean and standard deviation (`SD') of differences in four source parameters, calculated for sources that have been observed more than once.
Robust statistics have been used to calculate these numbers. We consider two populations: only those sources that emit at one RM, and all sources. The number of data points in each category is shown in the second row.
}
\label{parameter_comparison_multiple_obs.tab}
\begin{tabular}{lcccc}
\hline
Parameter					&	\multicolumn{2}{c}{Sources with one RM}	&	\multicolumn{2}{c}{All sources} \\
						& 	\multicolumn{2}{c}{(29 source pairs)}	&	\multicolumn{2}{c}{(111 source pairs)} \\
						&	Mean	&	SD		&	Mean	& SD		\\
\hline
$\Delta$ Stokes $I_\mathrm{ref}$ (mJy)	&	  -0.1		&	0.9		&	   0.1		&	33.8		\\
$\Delta$ Stokes $L_\mathrm{ref}$ (mJy)	&	   0.3		& 	0.9		&	   0.0		&	  1.6		\\
$\Delta$RM (\radm)					&	   0.1		& 	0.5		&	  -0.3		&	  3.1		\\
$\Delta\alpha_\mathrm{L}$			&	   0.0		&	0.1		&	   0.0		&	  0.3		\\
\hline
\end{tabular}
\end{table}

\begin{figure*}
\begin{centering}
\resizebox{0.32\hsize}{!}{\includegraphics[trim={0cm 0cm 0 0}, clip]{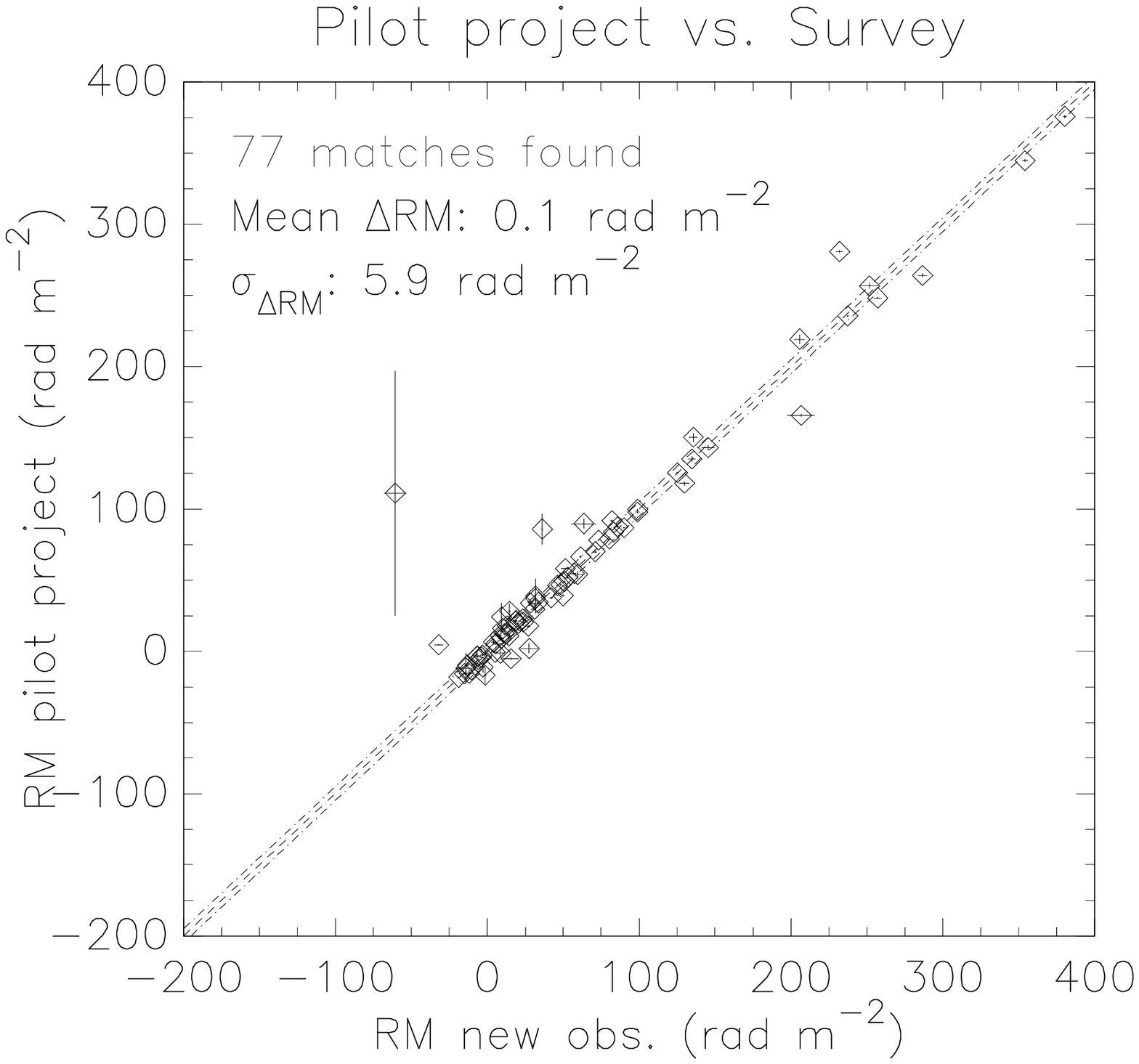}}
\resizebox{0.32\hsize}{!}{\includegraphics[trim={0cm 0cm 0 0}, clip]{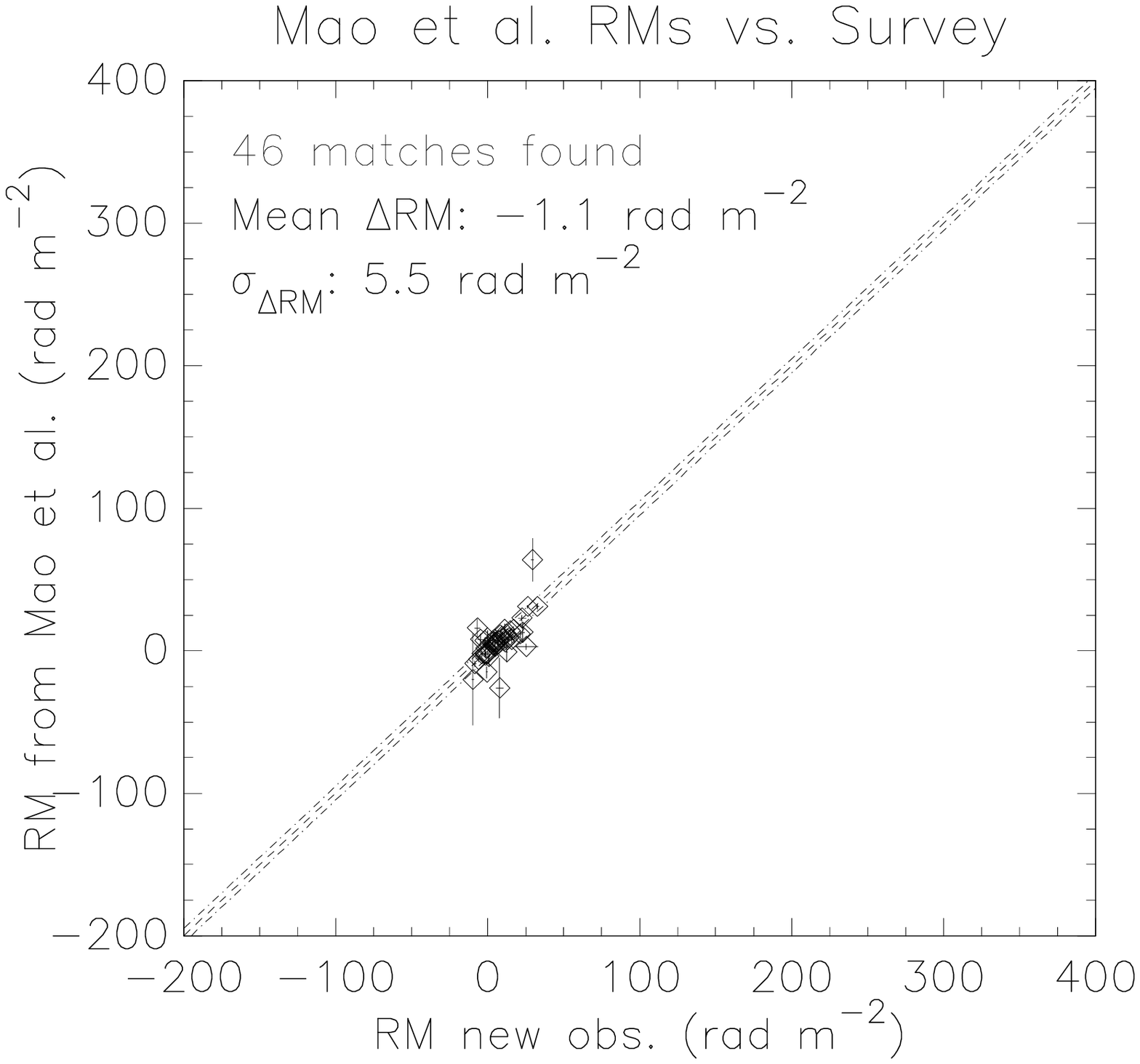}}
\resizebox{0.32\hsize}{!}{\includegraphics[trim={0cm 0cm 0 0}, clip]{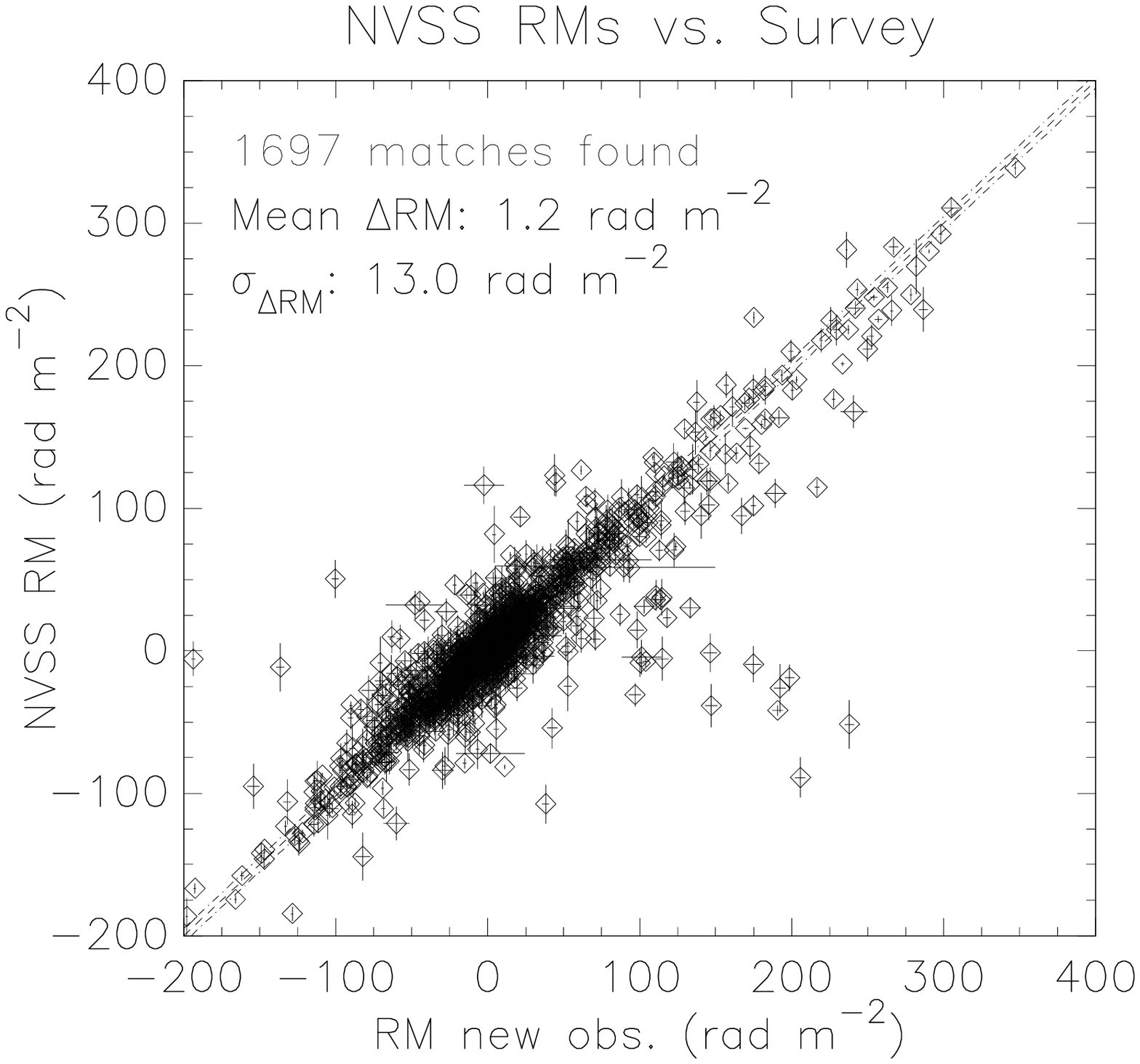}}
\caption{
Comparison between RMs from S-PASS/ATCA, the pilot project for S-PASS/ATCA described in Section~\ref{quality.sec} (panel on the left), RMs published towards the south Galactic Pole by \citet{mao2010} (middle panel), and from TSS09 (panel on the right).
Also shown is the one-to-one line, which is flanked by lines that are shifted in the $y$ direction by $\pm 5$~\radm . 
Each panel shows the number of sources with counterparts, and the mean and standard deviation of $\Delta$RM, the difference in RM between a source from S-PASS/ATCA and its counterpart.
}
\label{RM_comparison.fig}
\end{centering}
\end{figure*}

\begin{figure*}
\begin{centering}
\resizebox{0.32\hsize}{!}{\includegraphics[trim={0cm 0cm 0 0}, clip]{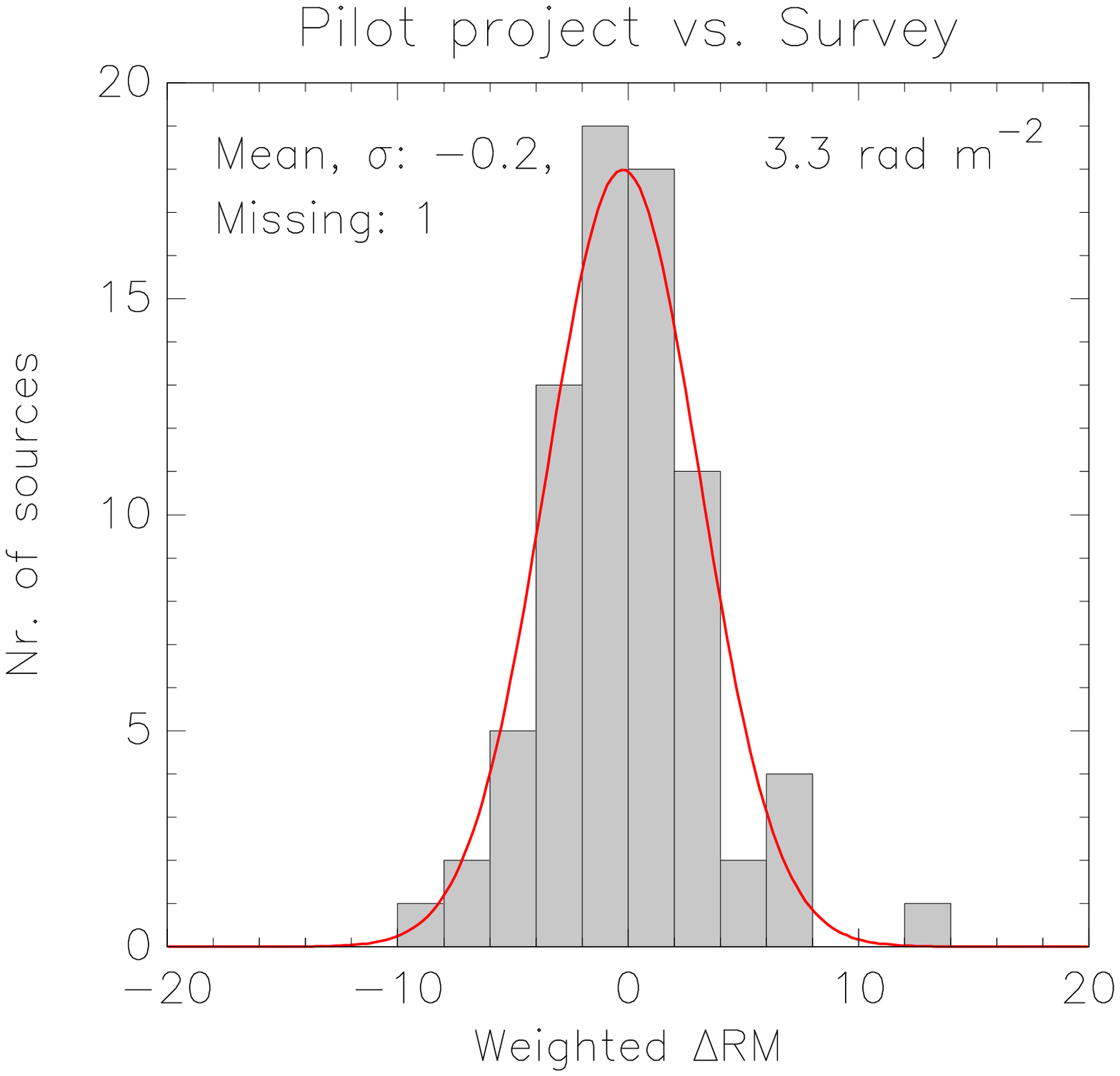}}
\resizebox{0.32\hsize}{!}{\includegraphics[trim={0cm 0cm 0 0}, clip]{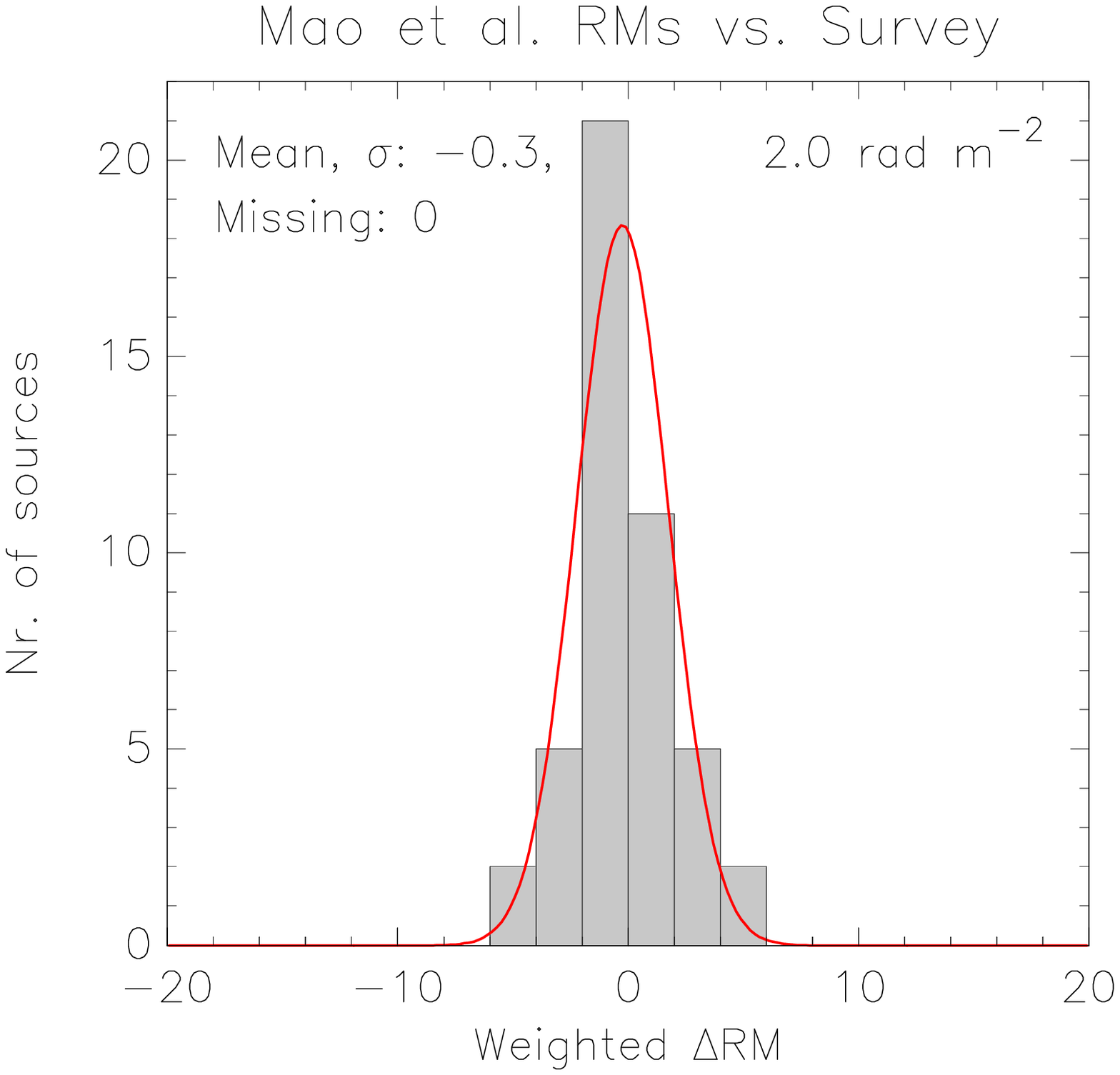}}
\resizebox{0.32\hsize}{!}{\includegraphics[trim={0cm 0cm 0 0}, clip]{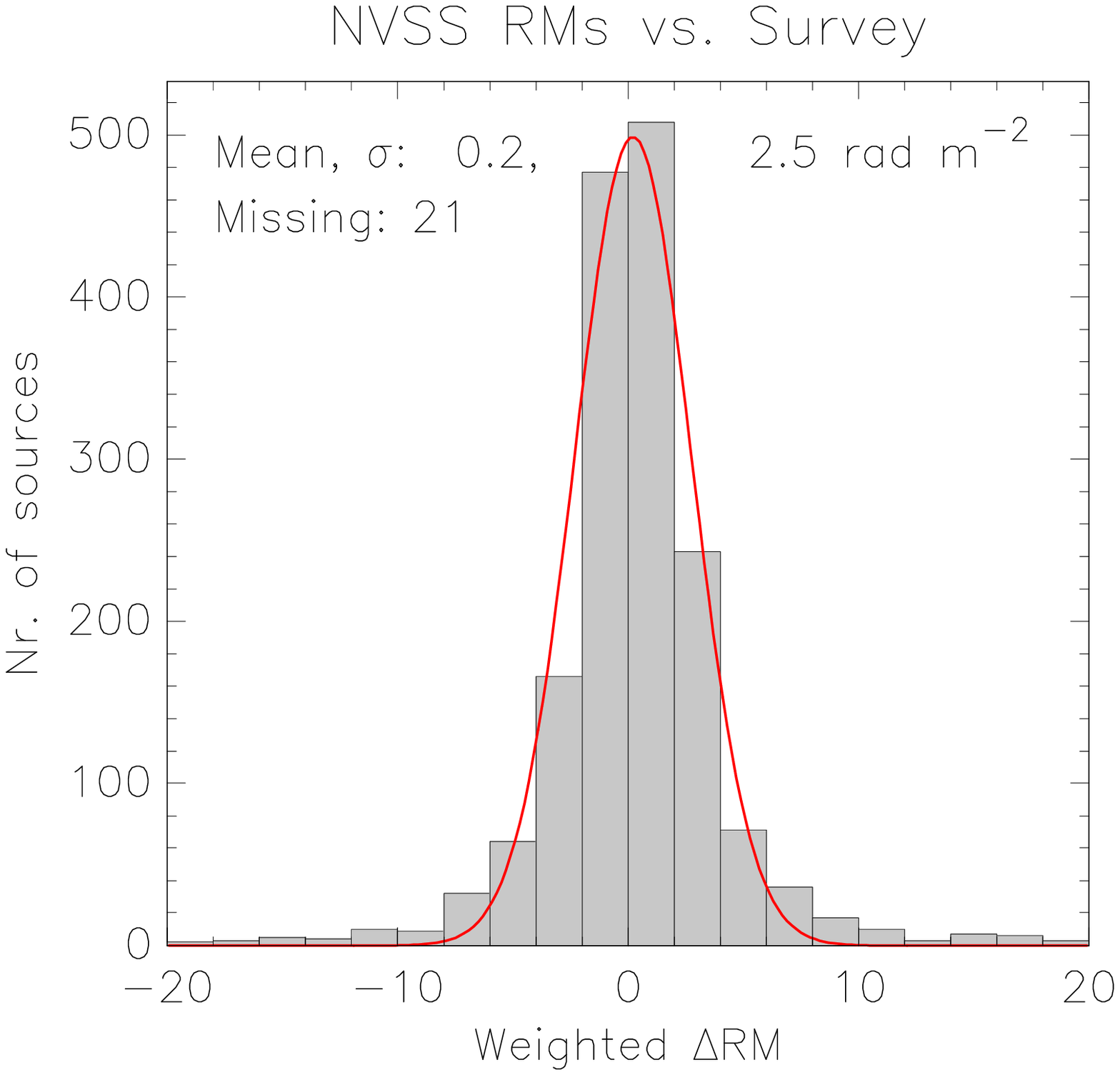}}
\caption{
Distributions of $\Delta$RM from Fig.~\ref{RM_comparison.fig} normalized by the uncertainty in each $\Delta$RM. This uncertainty is calculated by adding the uncertainties in RM from S-PASS/ATCA and its counterpart in one of the other catalogues in quadrature. 
Each panel shows the mean and standard deviation of the distribution of normalized $\Delta$RM, which are calculated using robust statistics. These numbers also determine the position of the centre and the width of the Gaussian distribution shown in red; the amplitude of this Gaussian is chosen such that the surface area under each curve is the same as the total number of data points in each histogram. The number following 'Missing' indicates the number of points that fall outside the plot range.
}
\label{RM_comparison_histograms.fig}
\end{centering}
\end{figure*}

In Figs.~(\ref{RM_comparison.fig}) and (\ref{RM_comparison_histograms.fig}) we compare the RMs from S-PASS/ATCA with the RMs of the pilot project, and the RM catalogues that have been published by \cite{mao2010} and TSS09. 
For each target field in S-PASS/ATCA or in the pilot project, we determined if the brightest or the second brightest source in Stokes $I$ produced the detection with the highest polarized signal-to-noise ratio $L_{1,\mathrm{ref}}/\sigma$: a visual inspection of the models that were fitted to data from our pilot project showed that the brightest source in Stokes $I$ does not necessarily emit the strongest polarized signal $L_{1,\mathrm{ref}}$.
Therefore, we selected the RM of source that was detected with the highest polarized signal-to-noise ratio for our comparison with the other catalogues. 
Two sources were removed from the pilot data set because they showed jumps in polarized flux density.
We cross-correlated RMs from S-PASS/ATCA with the RMs from the other catalogues, selecting in each case the nearest counterpart if that counterpart lies within 3\arcmin.
In Fig.~(\ref{RM_comparison.fig}) we plot the RMs from the various surveys against each other, while in Fig.~(\ref{RM_comparison_histograms.fig}) we show distributions of the RM difference between two catalogues, normalized using the combined uncertainty in RM.

The distributions of $\Delta$RM and the normalized $\Delta$RM have a mean close to zero, which implies that there is no strong bias in the RMs that we selected from S-PASS/ATCA. 
The scatter in the distribution of $\Delta$RM is much larger when we compare RMs from S-PASS/ATCA with RMs from Taylor et al., instead of RMs from Mao et al. or our pilot project.
However, this largely reflects the larger error bars for RMs from the catalogue by Taylor et al., as Fig.~(\ref{RM_comparison_histograms.fig}) shows. 
\cite{stil2011} suggested that the uncertainties in RM in the catalogue by Mao et al. are underestimated by a factor of about $\sqrt{2}$, and in the catalogue by Taylor et al. by a factor of 1.22. 
If we apply these correction factors, then the mean and standard deviation of the normalized $\Delta$RMs are -0.2 and 1.5 \radm, respectively, for the catalogue by Mao et al., and 0.2 and 2.1 \radm\, respectively, for the catalogue by Taylor et al.
This puts the RMs from S-PASS/ATCA and Mao et al. and their associated uncertainties in very good agreement.

\begin{figure}
\begin{centering}
\resizebox{0.9\hsize}{!}{\includegraphics[]{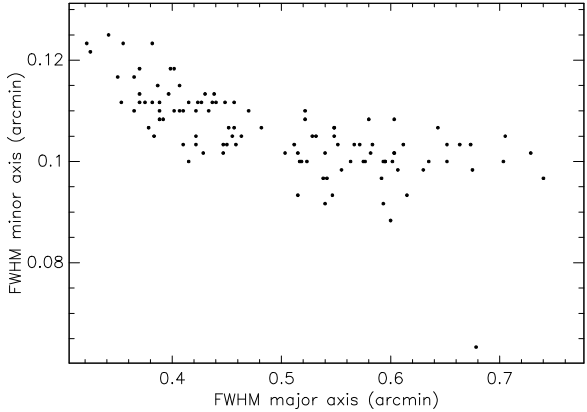}}
\caption{
Sizes of the restoring beam at 2.2 GHz used to create CLEANed images for targets from the pilot project. Note the difference in scale between the coordinate axes.
}
\label{beamsizes_pilot_project.fig}
\end{centering}
\end{figure}

The RMs from S-PASS/ATCA match the RMs from Taylor et al. and Mao et al. well, which indicates that most S-PASS/ATCA targets are dominated by polarized emission at a single RM. We return to this in Section~\ref{science.sec}.
Remarkably, the scatter in the distribution of normalized $\Delta$RM is largest when we compare RMs from S-PASS/ATCA with our own pilot project. Perhaps this can be explained to some degree by the higher angular resolution of the data from the pilot project (Fig.~\ref{beamsizes_pilot_project.fig}); the data from Mao et al. and Taylor et al. use beams with a FWHM of $\sim$ 30\arcsec\ and 45\arcsec, respectively.

\begin{figure}
\begin{centering}
\resizebox{0.85\hsize}{!}{\includegraphics[]{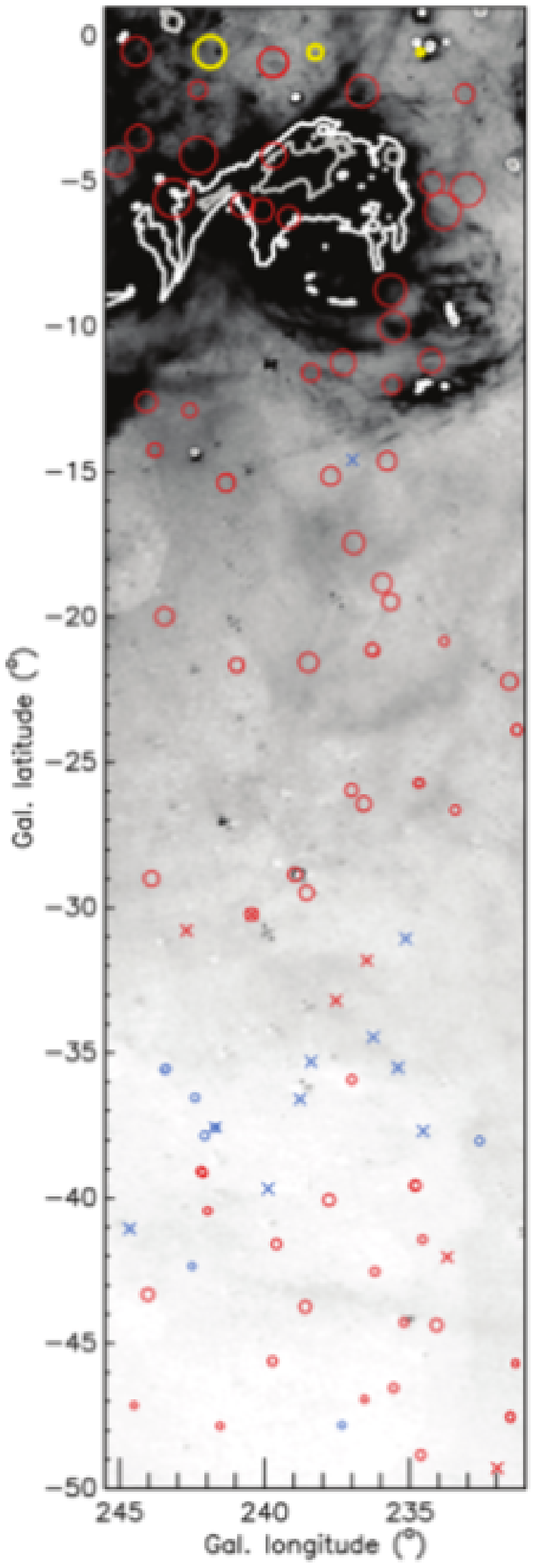}}
\caption{
RMs from the pilot project, on top of H$\alpha$ intensities from \citet{finkbeiner2003}. 
Positive RMs are shown in red, negative in blue. The yellow circles at the top of the figure indicate RMs of $\pm$ 250, 50, and 10~\radm. Crosses show $|$RM$|$ $<$ 10~\radm , otherwise these would be difficult to see.
The H$\alpha$ map shows intensities from 0 (white) - 25 (black) Rayleigh (not correcting for interstellar extinction); light and dark grey lines indicate contour levels of 50 and 100 Rayleigh. 
Sources shown in this figure also satisfy the criteria outlined at the beginning of Section~\ref{quality.sec}.
}
\label{rms_pilot.fig}
\end{centering}
\end{figure}

\begin{figure*}
\begin{centering}
\resizebox{0.925\hsize}{!}{\includegraphics[trim={0cm 0cm 0cm 0cm}, clip]{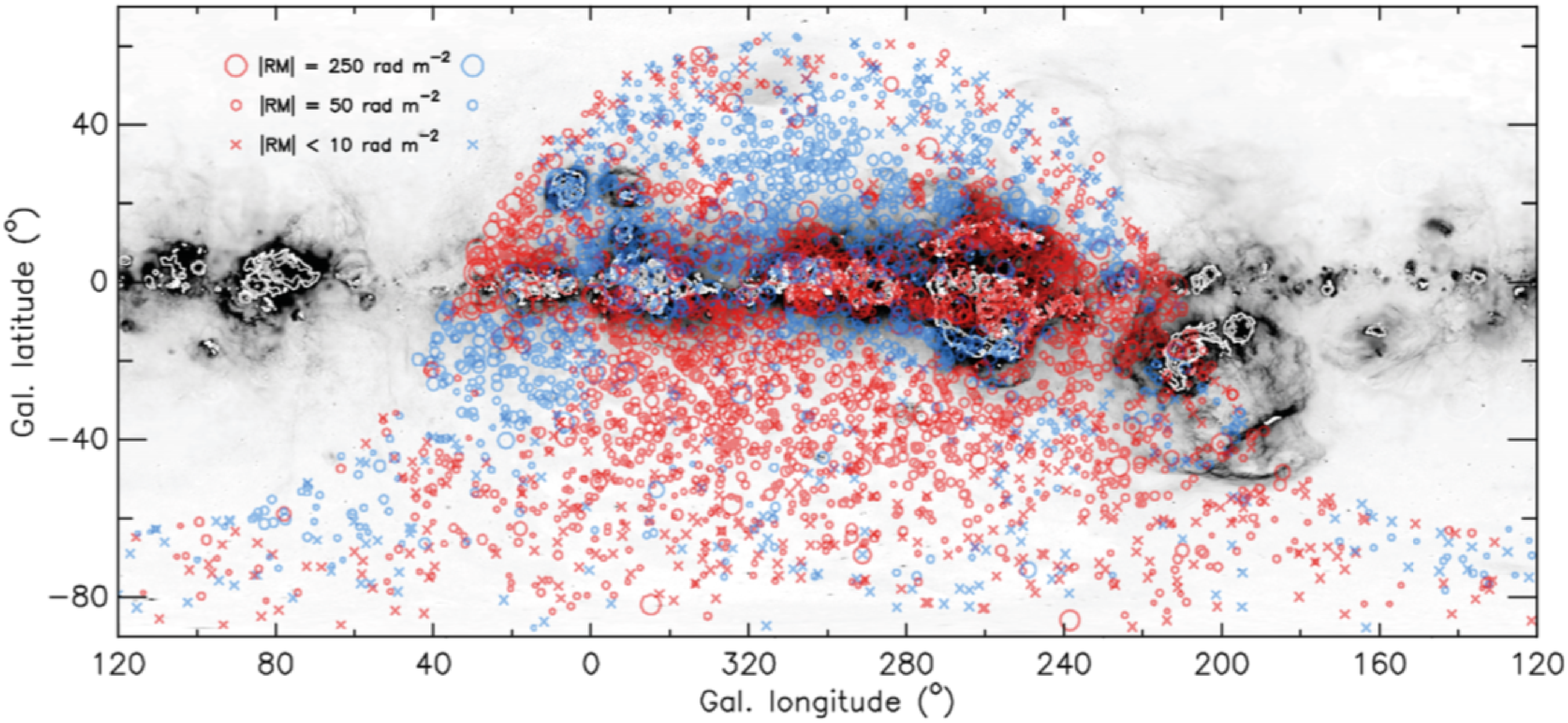}}
\resizebox{0.9\hsize}{!}{\includegraphics[trim={0cm 0cm 0cm 0cm}, clip]{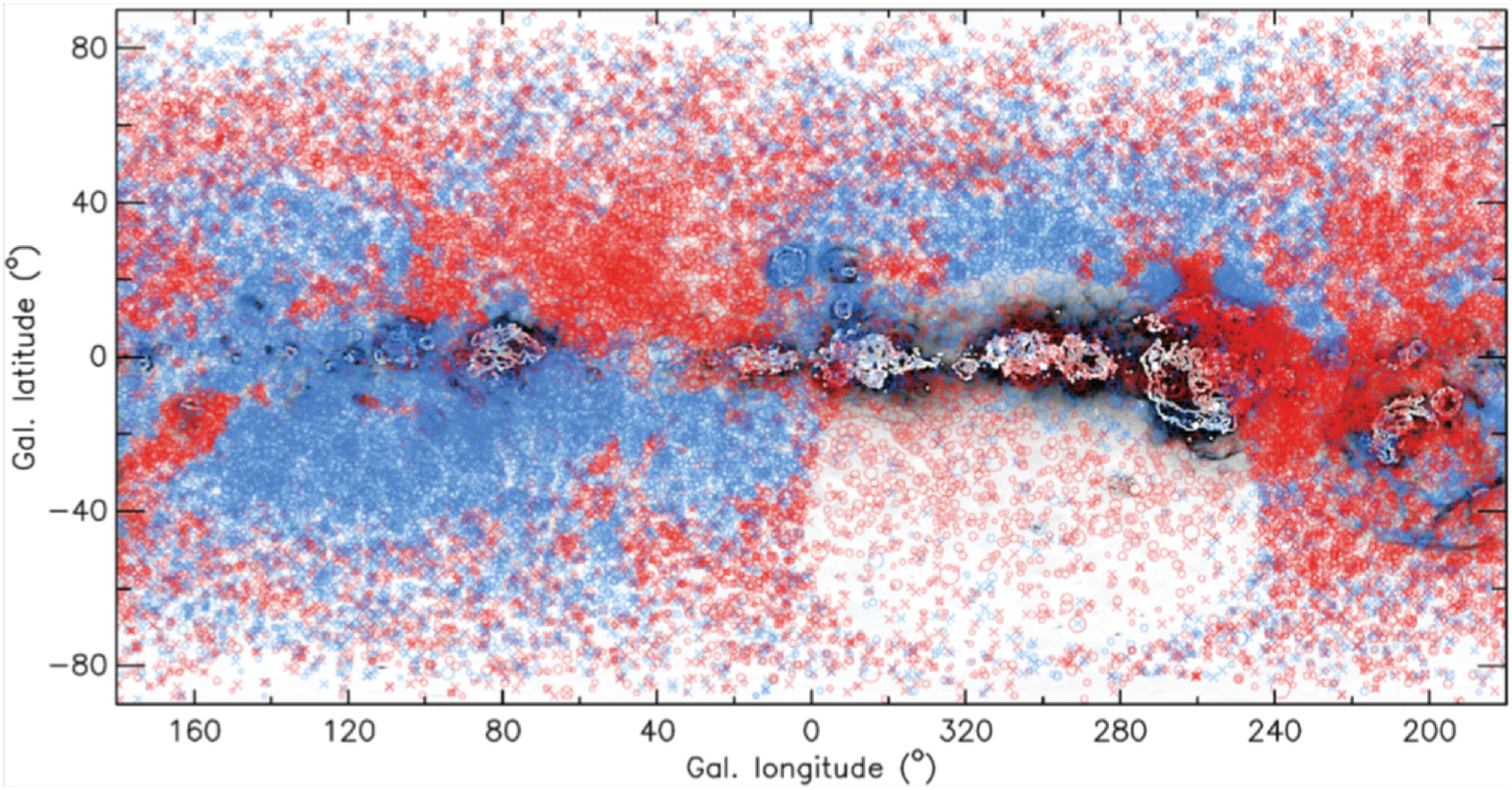}}
\caption{
\emph{Top:} map of RMs from S-PASS/ATCA that covers the southern sky.
\emph{Bottom:} all-sky map of RMs, combining data from S-PASS/ATCA and TSS09.
Symbols and colours are the same as in Fig.~(\ref{rms_pilot.fig}), symbol sizes and their corresponding RM values are shown in the top-left of the figure at the top. 
Note the different plot ranges that were used in these figures.
}
\label{sky.fig}
\end{centering}
\end{figure*}

\begin{figure}
\begin{centering}
\resizebox{\hsize}{!}{\includegraphics[trim={0cm 0cm 0cm 0cm}, clip]{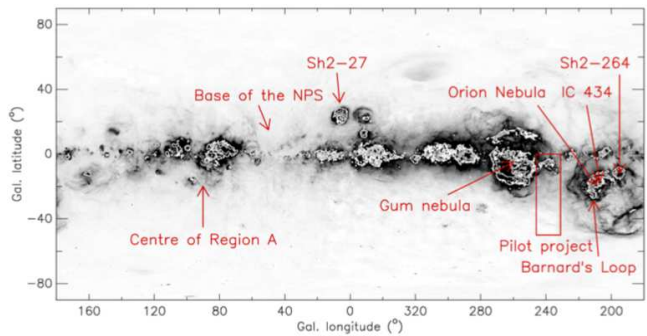}}
\caption{
Finder chart for regions that are mentioned in the text.
}
\label{finder_chart.fig}
\end{centering}
\end{figure}

\begin{figure*}
\begin{centering}
\resizebox{0.9\hsize}{!}{\includegraphics[trim={0cm 0cm 0cm 0cm}, clip]{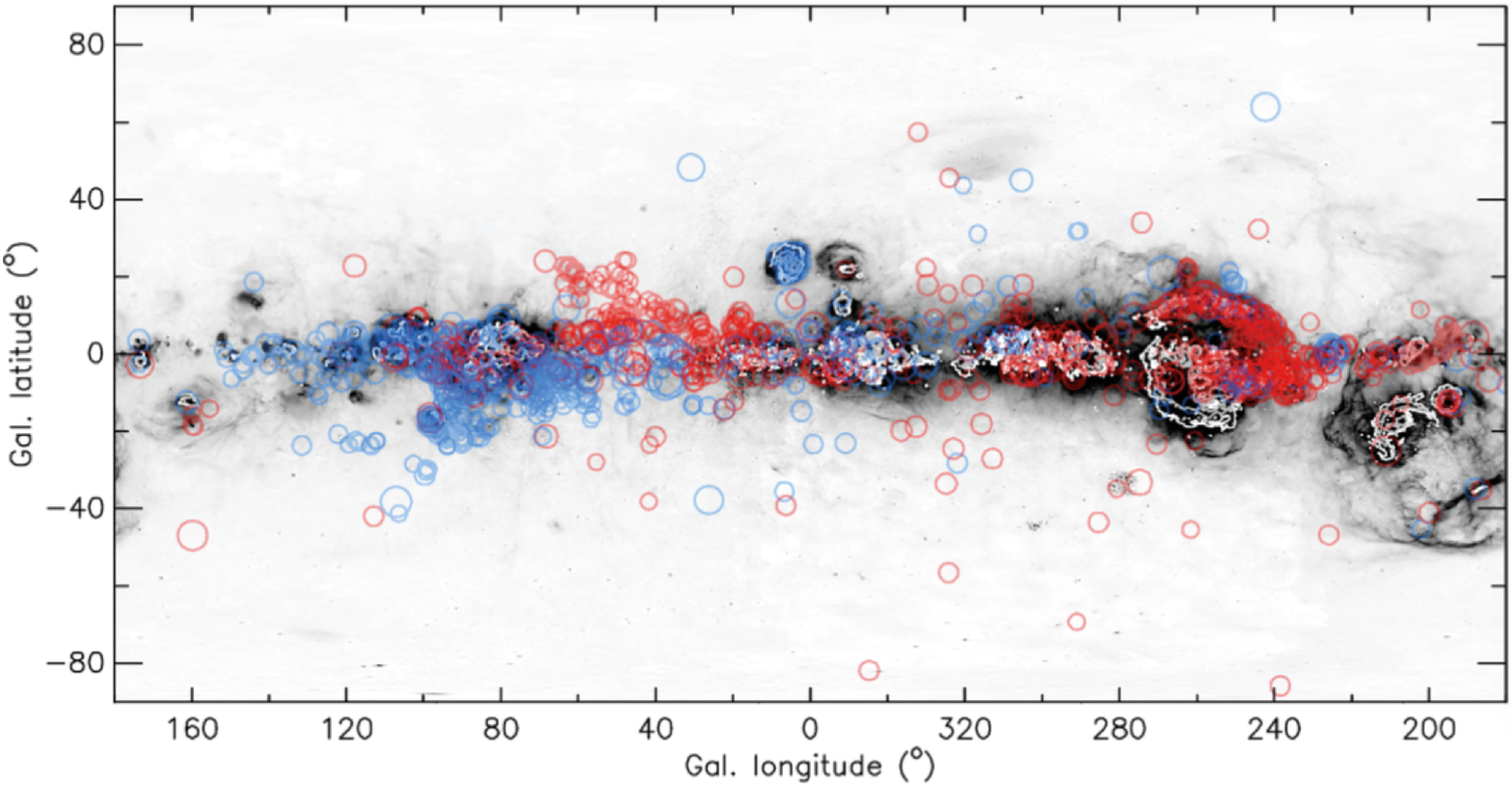}}
\caption{
Sources from TSS09 and S-PASS/ATCA that have $|$RM$|$ $>$ 150~\radm. Symbols and colours are the same as in Fig.~(\ref{sky.fig}).
}
\label{sky_large_rm.fig}
\end{centering}
\end{figure*}

\section{Results}\label{science.sec}
In this Section we present a preliminary scientific investigation of the S-PASS/ATCA data: first, an overview of Galactic science cases, then an analysis of the contribution by the Galactic foreground to the observed RMs, and finally, we investigate the ensemble properties of the extragalactic sources themselves.
3811 sightlines in the survey satisfy the selection criteria outlined in the first paragraph of Section~\ref{quality.sec} (and 116 sightlines from the pilot project), which translates as one source per five square degrees.

\subsection{Galactic science}
Fig.~(\ref{rms_pilot.fig}) shows RMs from the pilot project, plotted on top of H$\alpha$ intensities from \cite{finkbeiner2003}. Zooming out, this time plotting RMs from the S-PASS/ATCA survey, a striking pattern becomes visible where RMs are generally negative above the Galactic plane and generally positive below the plane (Figs.~\ref{sky.fig} and \ref{finder_chart.fig}). When adding RMs from TSS09, it becomes clear that more such regions exist, that the sign of RM changes as a function of longitude, and that RMs above and below the Galactic plane have the opposite sign in many of these regions. This butterfly pattern in RM has been associated with the large-scale magnetic field of the Milky Way since the earliest RM maps of the sky \citep{gardner1969}, and has been interpreted as a signature of an A0 dynamo operating in the Milky Way (e.g., \citealt{han1997}). S-PASS/ATCA makes this pattern more clearly visible in the southern sky.
Some of the regions with RMs of the same sign can be due to structures in the interstellar medium close to the Sun, so that they leave an imprint across a wide area on the sky. 
This could be the case for `Region A', an extended region with very negative RMs (see \citealt{simard-normandin1980}, according to whom Region A spans the region between 60\degr $< l <$ 140\degr\ and -40\degr $< b <$ 10\degr), and the North Polar Spur (NPS), a bright, polarized region that could be produced by the Galactic magnetic field wrapping around a nearby bubble of \hi\ gas (\citealt{wolleben2010}, \citealt{sun2015b}). 
Part of the NPS can be seen in RMs from S-PASS/ATCA as an enhancement in RM that starts around $l, b$ $\approx$ 20\degr, 5\degr. 

Fig.~(\ref{sky_large_rm.fig}) shows only sightlines with $|$RM$|$ $>$ 150~\radm . 
Sources with large $|$RM$|$s can be seen towards the Galactic Plane, or towards regions that are strong H$\alpha$ emitters like the Gum nebula (centred on $l, b$ $\approx$ 260\degr, 0\degr) and the \hii\ region surrounding $\zeta$~Oph, Sh2-27 ($l, b$ = 8\degr, 23\degr; RMs in this region have been analysed by \citealt{harvey-smith2011}). 
Two concentrations of large RMs can be seen at $l, b$ $\approx$ 90\degr,-20\degr\ and $l, b$ $\approx$ 60\degr, 20\degr : the first is probably related to Region~A.
Sources with $|$RM$|$ values much larger than a few hundred \radm\ are rare in other parts of the sky. They could have been detected with S-PASS/ATCA, and also with TSS09. However, in TSS09, such sources suffer from strong depolarization (see figure~1 in the paper by Taylor et al., \citealt{stil2007}, and \citealt{pasetto2016}) and they can be misinterpreted because of n$\pi$ ambiguities in polarization angle (see, e.g., \citealt{ma2017}).

\begin{figure*}
\begin{centering}
\resizebox{0.47\hsize}{!}{\includegraphics[trim={0cm 0cm 0cm 0cm}, clip]{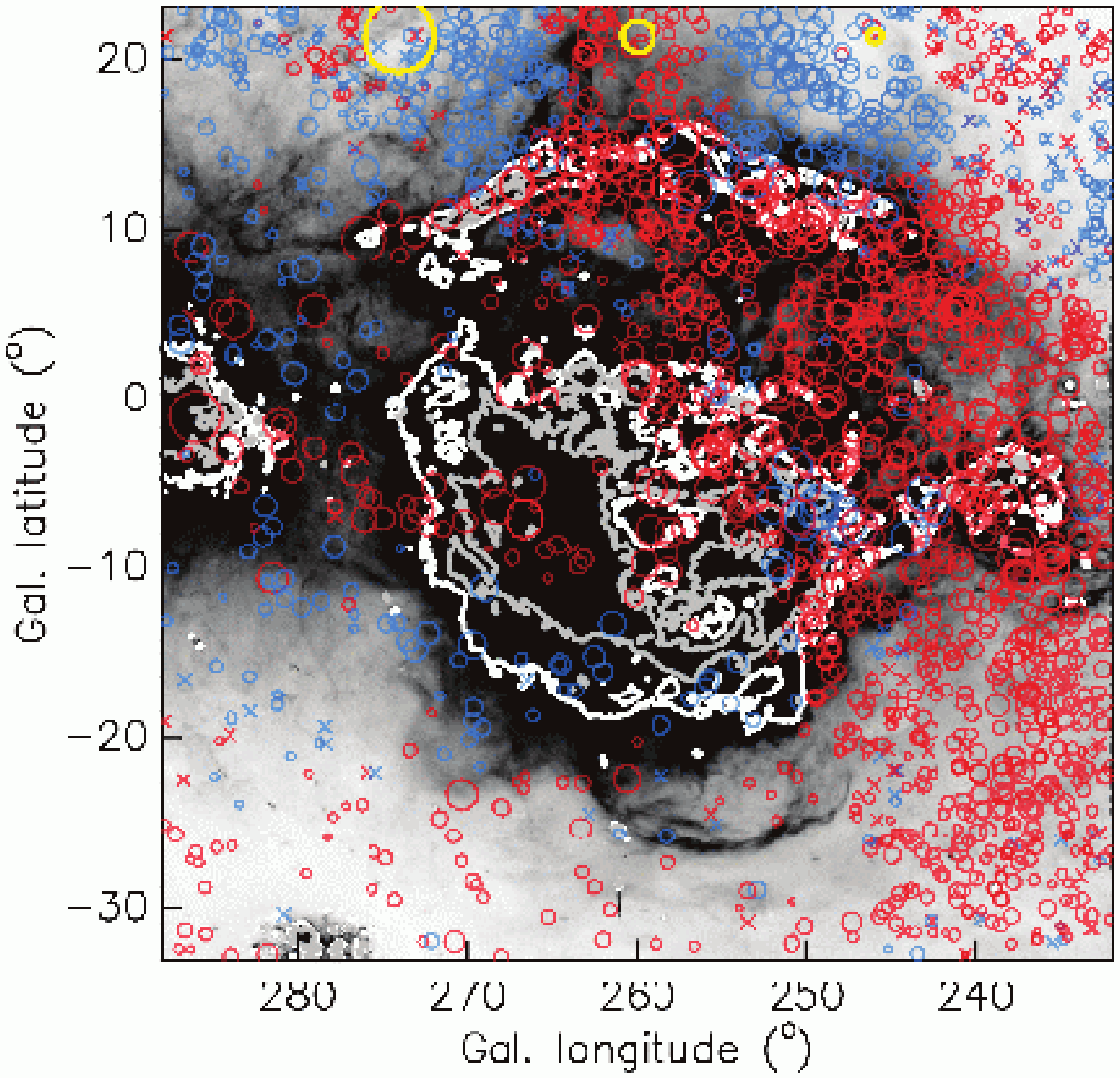}}
\resizebox{0.47\hsize}{!}{\includegraphics[trim={0cm 0cm 0cm 0cm}, clip]{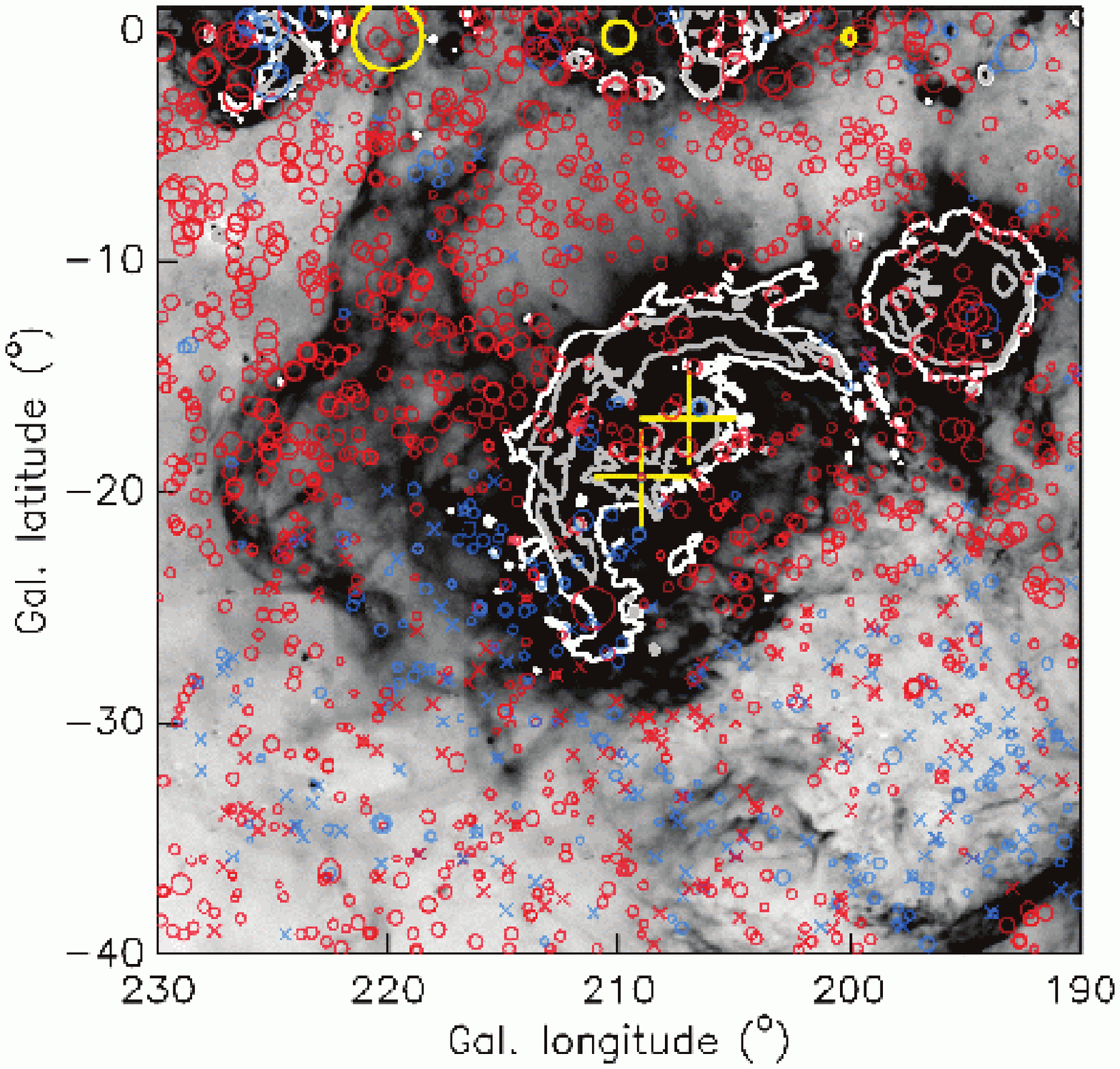}}
\caption{
RMs towards the Gum nebula (\emph{left}) and the Orion complex (\emph{right}). 
Both panels use RMs from S-PASS/ATCA and TSS09.
Symbols and colours are the same as in Fig.~(\ref{rms_pilot.fig}), except that now the legend shows $|$RM$|$s of 1250, 250, and 50~\radm. RMs from S-PASS/ATCA are drawn with thicker lines. In the panel on the right, the yellow pluses indicate the position of the Orion Nebula ($l, b$ = 209\degr,-19\degr) and the \hii\ region IC~434 ($l, b$ = 207\degr, -17\degr).
}
\label{RMs_Gum_Orion.fig}
\end{centering}
\end{figure*}

\begin{figure*}
\begin{centering}
\resizebox{0.45\hsize}{!}{\includegraphics[trim={0cm 0cm 0 0}, clip]{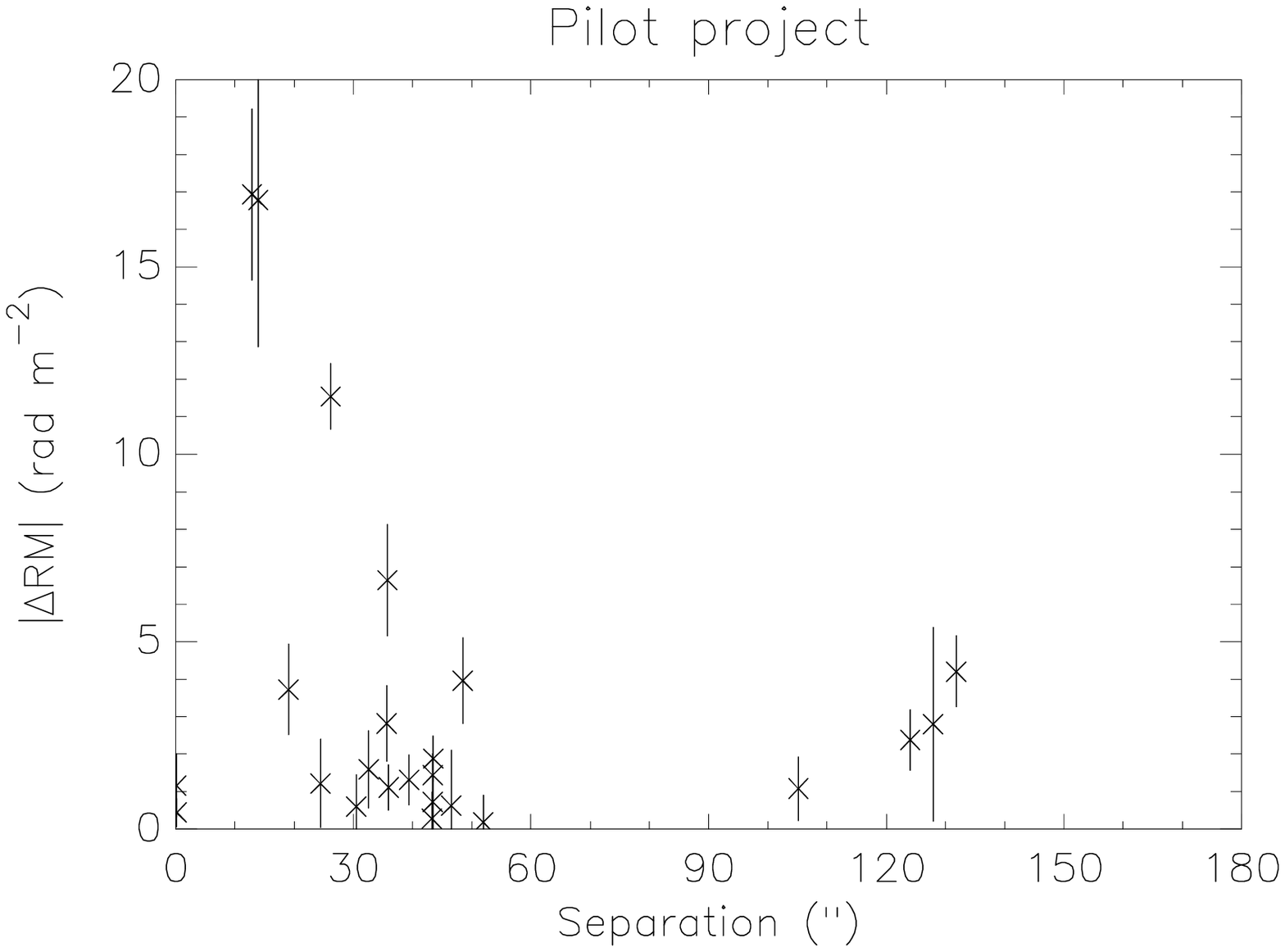}}
\resizebox{0.45\hsize}{!}{\includegraphics[trim={0cm 0cm 0 0}, clip]{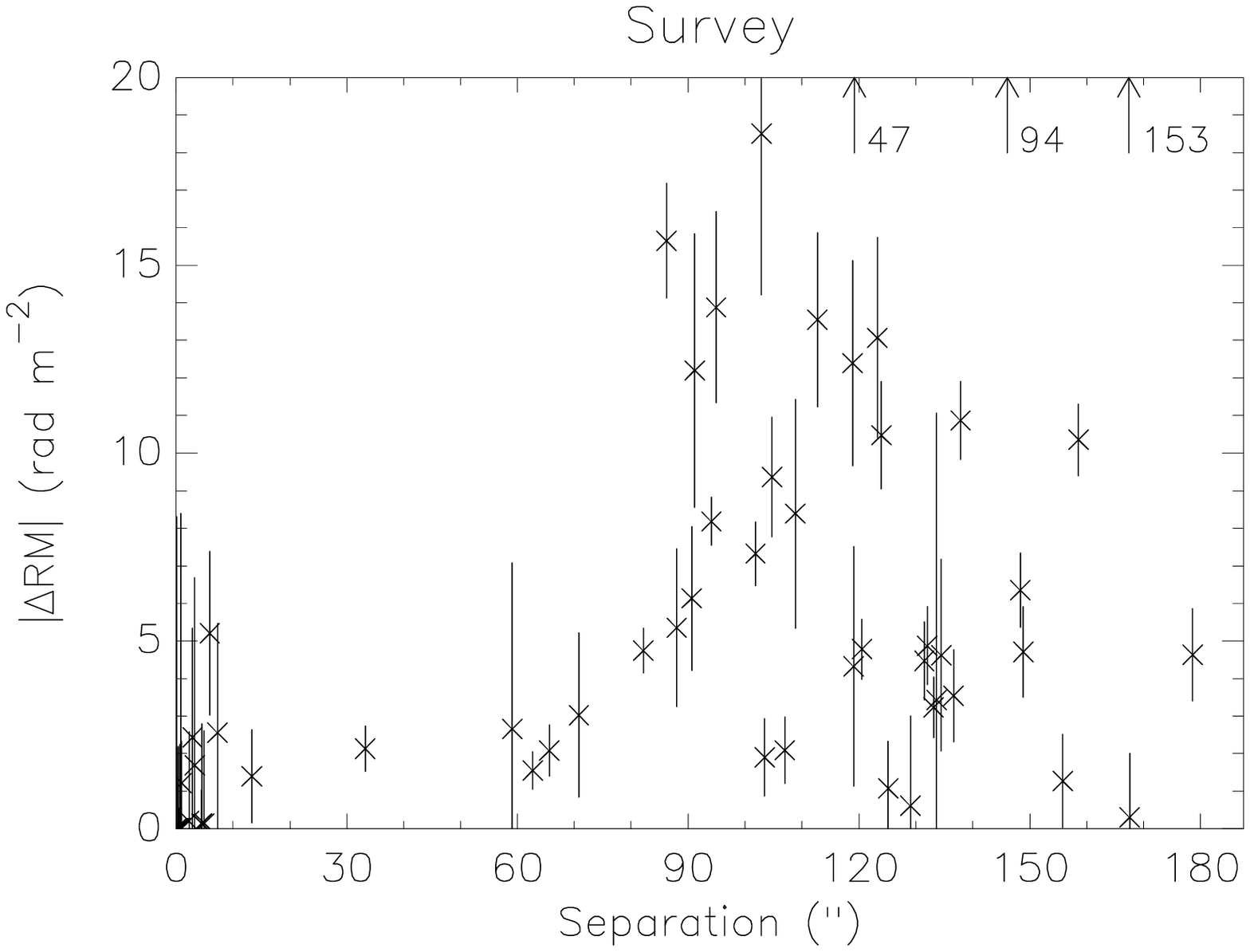}}
\resizebox{0.45\hsize}{!}{\includegraphics[trim={0cm 0cm 0 0}, clip]{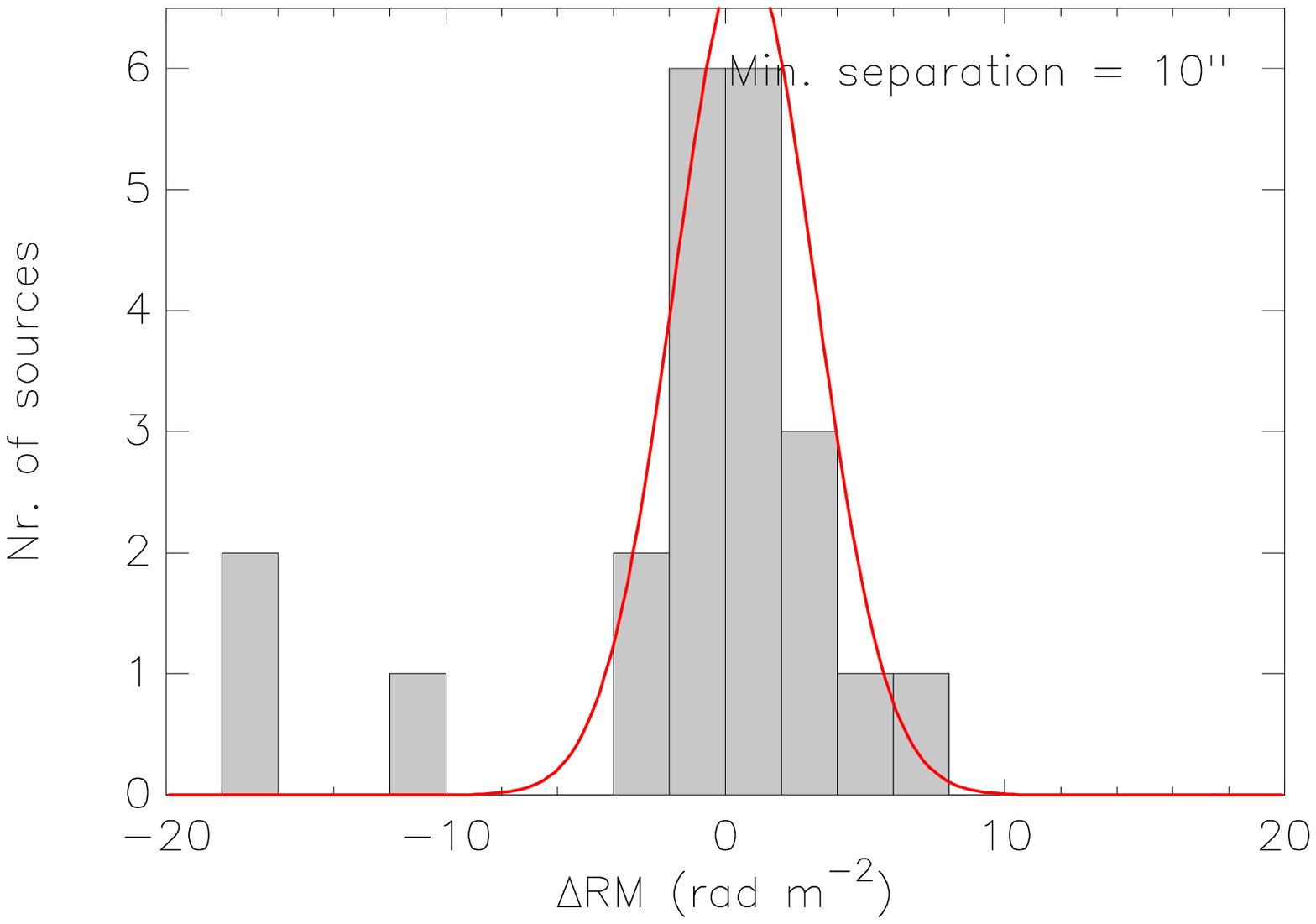}}
\resizebox{0.45\hsize}{!}{\includegraphics[trim={0cm 0cm 0 0}, clip]{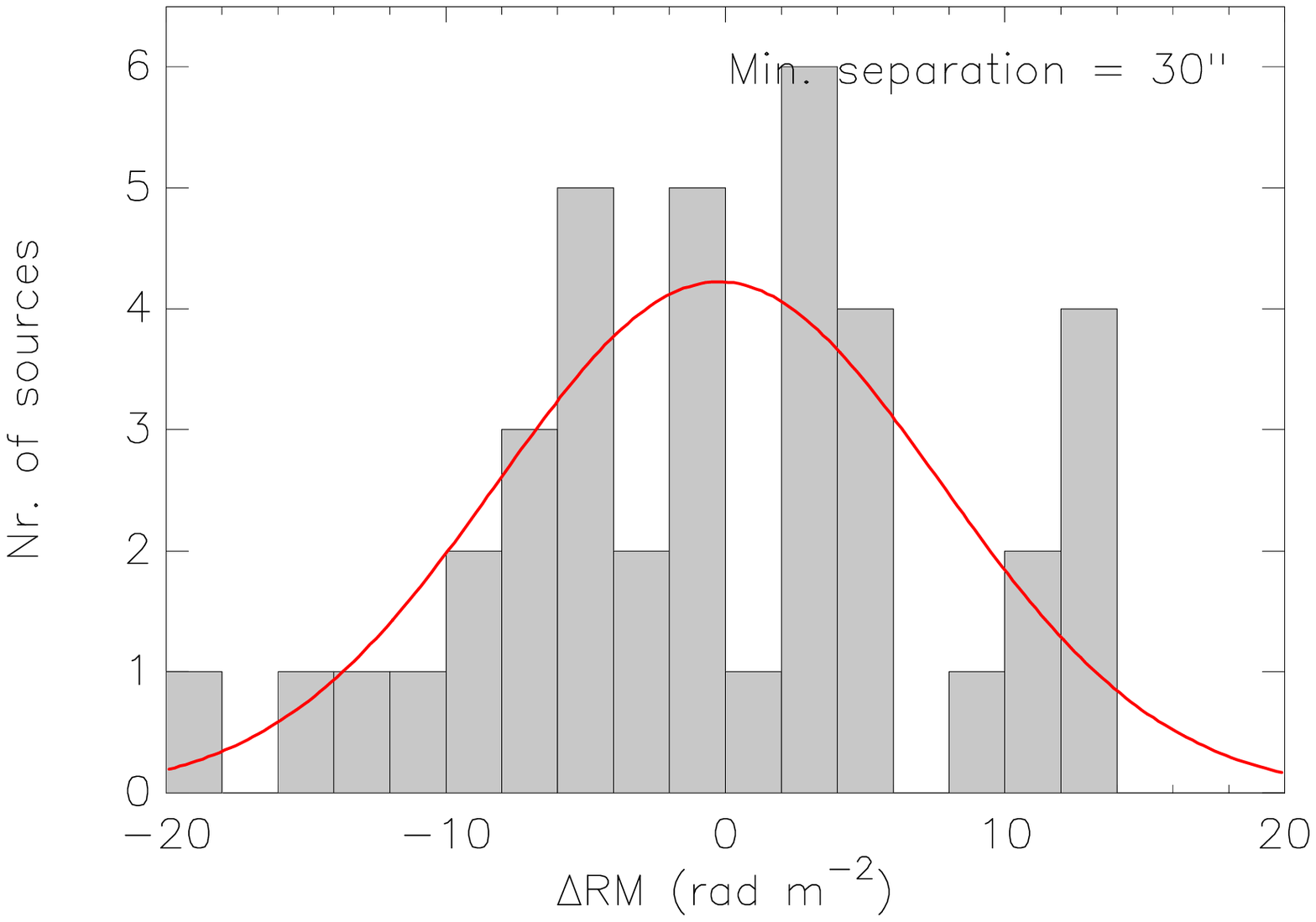}}
\caption{
Comparison of RMs measured for closely spaced sources. The top row of panels shows the RM difference, $\Delta$RM, between a pair of sources as a function of their separation, the bottom row of panels shows the distribution of $\Delta$RM together with a Gaussian that has the same mean and standard deviation as this distribution. 
The column on the left shows results from the pilot project, the column on the right for the full survey.
Three data points fall outside the plot range in the panel at the top-right; these have been indicated with arrows.
}
\label{deltaRM.fig}
\end{centering}
\end{figure*}

In Fig.~(\ref{RMs_Gum_Orion.fig}) we zoom in on two regions that are bright sources of H$\alpha$ emission: the Gum nebula and the Orion molecular cloud complex.
\cite{vallee1983} modelled the Gum nebula as a magnetic bubble close to the Sun, using 32~RMs that lie within 30\degr\ of the centre of the Gum nebula.
More recently, \cite{purcell2015} modelled the Gum nebula as an \hii\ region around a wind-blown bubble. 
RMs in the top part of the Gum nebula, which Purcell et al. used in their analysis, show a clear transition from inside the edge of the bubble (strong H$\alpha$ emission) to outside, over a very small distance (see fig.~3 in their paper). In the southern part of the Gum nebula, which is covered by S-PASS/ATCA but not by TSS09, RMs are much less correlated with regions that are bright in H$\alpha$. In fact, strips of RMs with the same sign intersect the white and grey contour lines in the south-western part of the Gum nebula. The part of the Orion molecular cloud complex that is outlined by the white and grey contour lines, Barnard's Loop, also does not show a strong correlation with RM. By contrast, RMs in the \hii\ region Sh2-264 (the $\lambda$ Orionis ring), centred on $l, b$ = 195\degr, -12\degr , increase in magnitude towards the peak in the H$\alpha$ intensity. \cite{harvey-smith2011} modelled the properties of the magnetic field in this \hii\ region.
It could be that inside the southern part of the Gum nebula and in Barnard's Loop the magnetic field is mostly perpendicular to the line of sight, or that the component $B_\|$, which contributes to RM, changes direction, perhaps on small scales. Then these regions leave no imprint on RM, and the RM that we measure would be built up along the rest of the line of sight.

\subsection{Extragalactic RM contribution}

The RMs of extragalactic sources increase in magnitude for sightlines closer to the Galactic plane, see, e.g., fig.~1 in \cite{schnitzeler2010}: the Milky Way contributes significantly to the observed RM values. 
One way for removing the contribution by the Milky Way to the observed RMs is to look for correlations between the RMs of nearby sources.
This technique has been applied successfully by, e.g., \cite{leahy1987}, \cite{schnitzeler2010}, \cite{pshirkov2013}, and \cite{oppermann2012, oppermann2015}, who found that extragalactic RMs have a standard deviation $\sigma$ $\approx$ 6~\radm , after subtracting the contribution by the Galactic foreground\footnote{The extragalactic RM contains all the RM contributions that originate beyond the Milky Way, and includes the RM that is accumulated inside the radio source itself.}.
\cite{xu2014b}, using a different analysis method, find that the residual RMs have $\sigma$ $\approx$ 13-15 \radm .
The bulk of the data points that were used in these analyses come from the RM catalogue by TSS09, and these RMs have large measurement uncertainties. The measurement uncertainties of the RMs in S-PASS/ATCA are much smaller, and are even smaller than the scatter in intrinsic RMs of extragalactic sources.
Here we apply a simpler technique to analyse the extragalactic RMs of radio sources in S-PASS/ATCA. Following \cite{conway1983}, \cite{simonetti1984, simonetti1986}, \cite{leahy1987}, and \cite{lazio1990}, we compare the difference in RM of pairs of radio sources separated by a small angular distance, in our case, up to 3\arcmin.
In this analysis, in addition to the selection criteria that we mentioned at the beginning of Section~\ref{quality.sec}, we impose the  additional requirements that sightlines have a reduced chi-squared of the fit to Stokes $I$ and $L$ of $<$ 2 and are fitted best by a single polarized component or by multiple components if their flux density ratio $L2/L1$ $<$ 1/3 (as determined at the reference frequency).

Fig.~(\ref{deltaRM.fig}) shows the distributions of $\Delta\mathrm{RM}$ and $|\Delta\mathrm{RM}|$ for sources from the pilot project and from the survey. We calculated the mean and standard deviation of the ensemble of $\Delta\mathrm{RM}$ using robust statistics, where we included source pairs that are separated by $>$ 10\arcsec\ (pilot project) and $>$ 30\arcsec\ (survey). 
Occasionally, the source finding algorithm identifies two components that are so close together that they probably correspond to peaks in the Stokes $I$ brightness distribution of a single source. The RMs of these sightlines would be highly correlated because of the size of the synthesized beam, not because of a physical effect. We excluded such pairs of sightlines from our analysis. Since the pilot project and the survey have different angular resolutions (Figs.~\ref{beamsizes.fig} and \ref{beamsizes_pilot_project.fig}), we use different minimum cut-off distances for these two data sets.
For the pilot project we find that the ensemble of $\Delta\mathrm{RM}$ has a mean = 1.0~\radm\ and $\sigma$ = 2.6~\radm , the distribution of $\Delta\mathrm{RM}$s from the survey has a mean = -0.2 \radm\ and $\sigma$ = 7.9 \radm\ (this distribution is clearly not Gaussian). 
These numbers are robust against changing the minimum separation of sources used in this analysis.
The values of $\sigma$ that we derived for the two distributions of $\Delta$RM are a factor of $\sqrt{2}$ larger than the width $\sigma$ of the extragalactic RM distribution: calculating the sum or the difference between two Gaussian distributions with the same $\sigma$ results in a Gaussian distribution that is wider by a factor of $\sqrt{2}$.
Therefore, we estimate that the distribution of extragalactic RMs has a width $\sigma$ = 1.9~\radm\ (pilot project), and 5.6~\radm\ (survey).
For comparison, the median measurement uncertainty in $\Delta$RM is 1.0~\radm\ for the pilot project, and 1.3 \radm\ for the survey. 
Since the value of $\sigma$ for the pilot project is not much larger than the median uncertainty of that sample, correcting for the measurement uncertainties would substantially reduce the width of the distribution of $\Delta$RM of the extragalactic sources. 

A visual inspection of the CLEANed Stokes $I$ maps of the pilot project and of the survey showed that the two RMs that we used to calculate $\Delta$RM belong to well-separated sources on the sky. Often these sources are of the double-lobed type, in a few cases these lobes are even resolved.
This inspection also identified six cases where data processing went awry. Since this is only a small number, we did not flag these cases in our analysis (but they are included in the data release).
To interpret our results, we consider two scenarios.
First, the two sightlines that we analysed could belong to different sources that are seen almost in the same direction purely by chance. In this case the intrinsic RMs of the two sources could be very different, which results in a wide distribution for $\Delta$RM. We observe a narrow distribution of $\Delta$RM, so, in this scenario, the sources must produce internally hardly any RM. 
Second, if the two sightlines belong to two components of the same physical source (for example, two radio lobes), then it is possible that the two components each produce very large $|$RM$|$s intrinsically, as long as the RM difference between the two components is small, to match our observations.
The pairs of sightlines that we analysed cover a wide range in physical separation between the two components, and the physical sources these components belong to cover a wide range in inclination with respect to the line of sight. We expect that this widens the distribution of $\Delta$RM, unless the sources produce intrinsically RMs that are close to zero.
We conclude that in both scenarios the intrinsic RMs of the extragalactic sources are probably very small.
Interestingly, \cite{athreya1998} reached the opposite conclusion in their sample of bright radio galaxies at $z$ $>$ 2. 
If extragalactic sources produce intrinsically RMs that are close to zero, then the Milky Way must be the dominant contributor to the observed RMs.
Furthermore, the Milky Way foreground has to be smooth in RM on such small angular scales; \cite{simonetti1984, simonetti1986} and \cite{leahy1987} reached the same conclusions. 
The small value for $\sigma$ that we calculated for source pairs selected from the pilot project implies that for the sources in this sample, the synchrotron-emitting and Faraday-rotating media inside and outside the radio lobes of these sources must be simple, otherwise we would have measured a wider range of RMs. 
Simple sources are not affected strongly by depolarization even at frequencies $\sim$ 100 MHz, allowing them to be detected with the Low-Frequency Array and the Murchison Widefield Array (\citealt{mulcahy2014}, \citealt{lenc2016}, \citealt{vaneck2017, vaneck2018}). In fact, \cite{vaneck2018} identified more than 2/3 of their sources as hotspots in FR II galaxies.
It is not clear why source pairs from the survey have a much wider RM distribution than source pairs from the pilot project.

\subsection{Intrinsic properties of extragalactic sources}\label{intrinsic_properties.sec}
\begin{figure}
\begin{centering}
\resizebox{0.8\hsize}{!}{\includegraphics[trim={0cm 0cm 0 0}, clip]{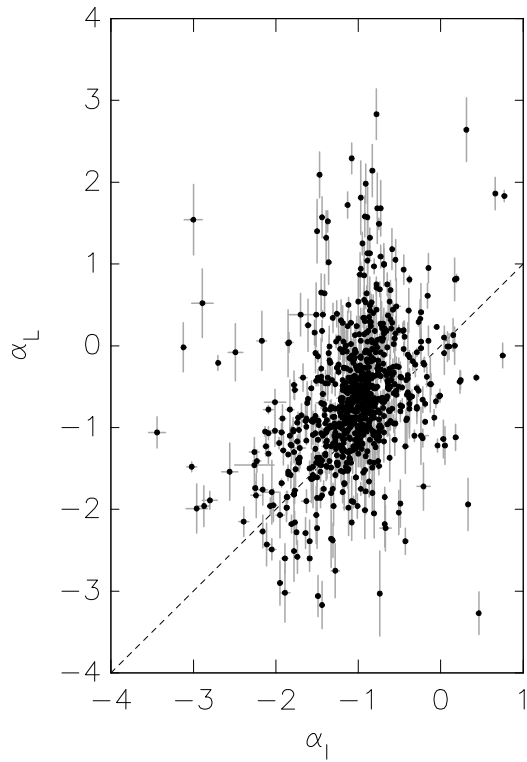}}
\caption{
Comparison between spectral indices fitted to Stokes $I$ (`$\alpha_I$') and $L$ (`$\alpha_L$').
The dashed line shows the one-to-one line.
Sightlines that are plotted in this figure emit at only one RM, satisfy the criteria outlined at the beginning of Section~\ref{quality.sec}, and the fits to Stokes $I$ and $L$ must have a reduced chi-squared $<$ 2.
}
\label{SI_distribution.fig}
\end{centering}
\end{figure}

In Fig.~(\ref{SI_distribution.fig}) we compare the fitted spectral index in Stokes $I$ with the fitted spectral index in $L$. 
We selected only sightlines for which the fits to Stokes $I$ and to $L$ both have a reduced $\chi^2$ $<$ 2.
Furthermore, the sources that are shown in this figure emit at a single RM, therefore, one would expect $\alpha_\mathrm{I}$ to be equal to $\alpha_\mathrm{L}$ to within the measurement uncertainties.
However, this is not the case: a large number of sources is fitted even with a positive $\alpha_\mathrm{L}$.  
We checked that source confusion in Stokes $I$ is not the main cause for this difference between the spectral indices.
In these cases, dividing Stokes $Q$ and $U$ by Stokes $I$ to remove spectral index effects produces erroneous results. In RM synthesis, one considers only sources that emit at a single RM, and in the past the ratios of flux densities $Q/I$ and $U/I$ have been used as input to RM synthesis.
Fig.~(\ref{SI_distribution.fig}) shows that this is not the right approach, and we discourage its usage. In \cite{schnitzeler2017} and S18 we provided mathematical arguments against this approach.

We investigated which fraction of sources that we analysed shows emission at more than one RM: answering this question will tell us more about the intrinsic complexity of the radio sources themselves, and also whether RMs that were determined from narrow-band observations like TSS09 are reliable.
If a large fraction of sources contains more than one component, then those components can show complex interference patterns in Stokes $Q$ and $U$ as a function of frequency. In that case, if the RM is determined by calculating the derivative of the polarization angle with respect to wavelength squared, then narrow-band observations at different frequencies would yield different RMs.

\begin{figure}
\begin{centering}
\resizebox{0.9\hsize}{!}{\includegraphics[trim={0cm 0cm 0 0}, clip]{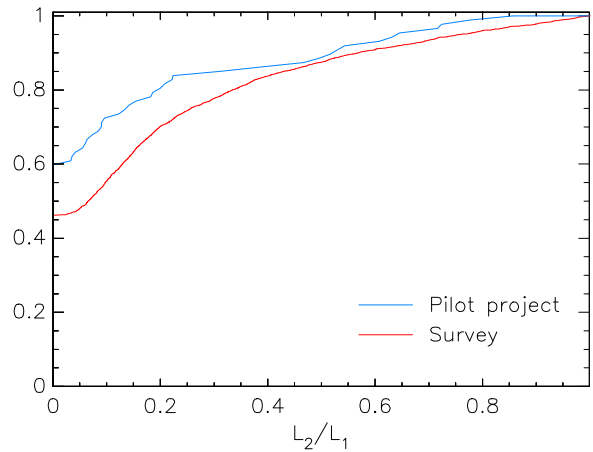}}
\caption{
Cumulative distribution of the ratio of flux densities $L_{2,\mathrm{ref}}/L_{1,\mathrm{ref}}$ for targets from the pilot project (blue line) and the survey (red line).
Sometimes the fitting procedure re-arranges the order of the components, in which case the component labeled `2' becomes the brightest polarized component. In those cases we swap $L_{1,\mathrm{ref}}$ and $L_{2,\mathrm{ref}}$, so that the ratio between these two flux densities is again $<$ 1.
}
\label{L2L1_cumulative.fig}
\end{centering}
\end{figure}

Fig.~(\ref{L2L1_cumulative.fig}) shows that about 45-60 per cent of the sources in our sample emit at a single RM, and for about 80 per cent of sources the brightest polarized source component is more than three times as bright as the second brightest component.
\cite{anderson2015} and \cite{osullivan2017} analysed the prevalence of sources that emit at more than one RM, using observations at frequencies similar to those covered by S-PASS/ATCA. 
\cite{ma2019} investigated this using data from the Very Large Array between 1.15-2 GHz.
\cite{anderson2015} find that at most 88 per cent of the sources in their sample show simple RM spectra (see section 6.1 in their paper). \cite{osullivan2017} find that 37 per cent of the sources in their sample are fitted best with a single component, which means a single function of the type described by equation~(2) in their paper. In our case, if a source is fitted by a single component this means that the source emits at only one RM.
Ma et al. find that about 60 - 70 per cent of the polarized sources they analysed emit at one RM.
Neither the sample we use, nor the samples that were used by Anderson et al., O'Sullivan et al., and Ma et al., are complete in a statistical sense. Therefore, these numbers might change once statistically complete samples become available.

The high percentage of sources that emit at a single RM means that most of the RMs that were determined in the past using narrow-band observations can still be used safely. Furthermore, it simplifies analyses that try to break down extragalactic RM contributions into contributions by the sources themselves, intervening objects, and the cosmic web (e.g., \citealt{banfield2014}, \citealt{vacca2016}, \citealt{basu2018}).
By going to lower frequencies, it becomes possible to resolve polarized emission over narrower ranges in RM (equation~14 in S18), and the fraction of sources that emits at a single RM would drop.

Finally, we point out that 42 sources have a polarized flux density of more than 100 mJy at 2.1 GHz and are unresolved in CLEANed Stokes $I$ maps. These sources can serve as references for future projects with the ATCA and other telescopes in the southern hemisphere, including the Square Kilometre Array and its pathfinders.

\section{Data access	}\label{access.sec}
The source catalogue can be accessed on VizieR\footnote{We're working on uploading the catalogue to VizieR. For now, the latest version of the catalogue can be downloaded by following this link: https://bit.ly/2E7jr3W.}. This catalogue provides information on the polarization properties of the brightest and second brightest source in Stokes~$I$ in the field of view. For each of these two sightlines we list the properties of the brightest and second brightest polarized source components that we fitted to the data.
In addition, we publish the following data products on this website\footnote{https://atoa.atnf.csiro.au/S-PASS-ATCA.jsp}:
\begin{enumerate}
\item Calibrated uvfits data files,
\item CLEANed Stokes $I$ FITS images from the self-calibration pipeline,
\item Diagnostic plots like Fig.~(\ref{example_los.fig}) for each source in the catalogue, and
\item Results for all the models that we fitted to each source using $\textsc{Firestarter}$.
\end{enumerate}

\section{Summary}\label{conclusions.sec}
We presented S-PASS/ATCA, a survey of polarized sources in the southern hemisphere that we observed with the ATCA in the 16 cm band. 
Candidates were selected from the S-PASS survey conducted with the Parkes radio telescope. Our observations are able to detect sources with $|$RM$|$s of up to at least 1000~\radm .
Our sample contains more than 3800 sightlines, and has an angular resolution of $\sim$ 1-2\arcmin: about the same as the number of stars that can be seen by the naked eye on a clear night, and the angular resolution of the human eye, respectively.
The targets were observed in less than 90 hours in total, using a new observing strategy for the ATCA that we developed. We also developed a new calibration strategy, because the observations did not re-visit calibrator sources (with few exceptions). We analysed the data using the $\textsc{Firestarter}$ algorithm, a QU-fitting based algorithm that was published by S18.
The RMs that we derived from the survey show good agreement with RMs that have been published previously by TSS09, \cite{mao2010}, and a pilot project where we observed and calibrated S-PASS sources using standard procedures.

We show two RM maps of the Milky Way, one contains only sources from S-PASS/ATCA, the other is an all-sky map of RMs that combines S-PASS/ATCA and TSS09. Sources with large $|$RM$|$ are found mostly close to the Galactic Plane and towards bright \hii\ regions, and are rare elsewhere. We show a panoramic RM map of the Gum nebula; RMs published by TSS09 do not reach declinations that are far enough south to cover the Gum nebula in its entirety. Although RMs near the top of the Gum nebula are clearly different inside and outside the nebula, in the southern part of the Gum nebula no clear correlation is seen between the magnitude of RM and H$\alpha$ intensity. The part of the Orion molecular cloud complex that is brightest in \hii\ , Barnard's Loop, also does not show a clear correlation with RM.
Perhaps in these regions the magnetic field is mostly perpendicular to the line of sight, or the component $B_\|$ changes sign along the line sight, so that these regions do not leave an imprint on RM.

Some of our target fields contain more than one source in Stokes $I$, and we used the RMs that we measured in these fields to confirm that probably the Galactic contribution to the observed RMs is much larger than the intrinsic RMs of most polarized sources. Also, we concluded that the Faraday-rotating and synchrotron-emitting media in such sources must be distributed in a simple way, otherwise we would expect much larger RM differences.

For sources that emit at only one RM, we find that the spectral index that is fitted to Stokes $I$ is often (very) different from the spectral index that is fitted to $L$. In RM synthesis one often assumes that the two are the same, so that spectral index effects can be removed by analysing ratios of flux densities $Q/I$ and $U/I$. Because the two spectral indices are often different, calculating RM spectra from these ratios of flux densities leads to erroneous results.

About half the sightlines that we fitted can be described by emission at only a single RM, and in more than 8/10 cases the source that is brightest in $L$ is at least three times as bright as the source that is second brightest. This implies that most sightlines are dominated by emission at a single RM. Therefore, most of the RMs that have been derived in the past based on narrow-band observations at these frequencies are still safe to use. Observations that extend to lower frequencies might be able to resolve emission into discrete peaks or narrow continuous distributions.

\section*{Acknowledgements}
We thank the people at ATNF/CSIRO, and in particular those who live on the Narrabri site, for their commitment and enthusiasm. Furthermore, we thank Phil Edwards (ATNF/CSIRO) for his help with scheduling our observations, Jamie Stevens (ATNF/CSIRO) and Lister Staveley-Smith (International Centre for Radio Astronomy Research at The University of Western Australia) for their contributions to this project, Sui Ann Mao and Aristeidis Noutsos (both at the Max Planck Institute for Radio Astronomy) for her feedback on the manuscript and for his help with creating the figures, respectively, and Emil Lenc (University of Sydney, now ATNF/CSIRO) for sharing his ATCA map of PKS B1934-638 and its surroundings. We thank the anonymous referee for their constructive comments.
This work has been carried out in the framework of the S-band All Sky Survey collaboration (S-PASS).
The ATCA and Parkes radio telescopes are part of the Australia Telescope National Facility which is funded by the Commonwealth of Australia for operation as a National Facility managed by CSIRO. The Dunlap Institute is funded through an endowment established by the David Dunlap family and the University of Toronto. B.M.G. acknowledges the support of the Natural Sciences and Engineering Research Council of Canada (NSERC) through grant RGPIN-2015-05948, and of the Canada Research Chairs program.

\bibliography{ne6e}

\appendix
\section{Data calibration}\label{Appendix_A.sec}
We self-calibrate a target field using the $\textsc{miriad}$ task $\textsc{selfcal}$, but only if the field contains at least one source brighter than 200~mJy in Stokes~$I$.
The self-calibration process creates images that measure 1024 pixels on each side, with pixels that are 8\arcsec\ in size, using robust = 0.5.
We apply the $\textsc{miriad}$ algorithm $\textsc{mfclean}$ to create a list of CLEAN components for the central quarter of each image. This algorithm uses a gain of 0.1, and stops when the first negative component is encountered, if the absolute maximum residual falls below 0.02, or after 500 iterations, whichever comes first.
The self-calibration process consists of a number of iterations, listed in Table~\ref{selfcal.tab}.
Each iteration improves the calibration solution by including fainter sources when calculating the calibration solution, by increasing the number of subbands, or, in the case of calibrator sources, by changing from a phase-only self-calibration to an amplitude+phase self-calibration.
For calibrators, we specify the coefficients of a cubic polynomial fitted to the Stokes~$I$ spectrum of the source when running $\textsc{selfcal}$; these coefficients are generated by the task $\textsc{uvfmeas}$.
For all other targets, we use a list of CLEAN components in combination with multi-frequency synthesis to calculate selfcal solutions ($\textsc{selfcal}$ option `mfs'). If the band is split into multiple subbands, we use multi-model multi-frequency synthesis, with one CLEAN component model for each subband ($\textsc{selfcal}$ option `mmfs').
Because $\textsc{miriad}$ cannot merge two calibration tables that were derived for different numbers of subbands, selfcal creates two files, `Source N.cal' (contains the selfcal solution for the frequency band in its entirety) and `Source N.cal2' (contains the selfcal solution where the band has been split into multiple subbands).
Therefore, each field can have up to three data files associated with it: `Source N' (the original file, containing the calibration solution from PKS B1934-638), `Source N.cal', and `Source N.cal2'.

\begin{table}
\centering
\caption{
Overview of parameters used to self-calibrate the data. We list the number of subbands used, whether a phase-only or an amplitude+phase selfcal was used, and the minimum flux density in Stokes~$I$ of sources on which the selfcal solution is based. Only fields where at least one source has $I~>~200$~mJy are self-calibrated.
}
\label{selfcal.tab}
\begin{tabular}{lcccc}
\hline
Source type	&	Iteration	&	nfbin$^1$	&	selfcal type$^2$	&	clip level \\
\hline
Calibrators	&	1		& 	1	&	p		&	$^3$\\
			&	2		&	1	&	p		&	$^4$\\
			&	3		&	1	&  	a+p		&	$^4$\\
			&	4		&	4	&	p		& 	$^4$\\
Targets		&	1		&	1	&	p		&	$^3$\\
			&	2		&	1	&	p		&	$^4$\\
			&	3		&	2	&	p		&	$^4$\\
\hline
\end{tabular}
\begin{list}{}{}
\item[$^1$] `nfbin' specifies the number of subbands in $\textsc{miriad}$.
\item[$^2$] `p' means phase-only self-calibration, `a+p' means amplitude + phase self-calibration.
\item[$^3$] clip level is equal to $\max\left[\left(I_1 + I_2\right)/2 , 50~\mathrm{mJy}\right]$, where $I_1$, $I_2$ are the Stokes~$I$ flux densities of the brightest and second brightest source in the field of view, respectively.
\item[$^4$] clip level is $I_1$/3.
\end{list}
\end{table}

\begin{figure}
\begin{centering}
\resizebox{0.8\hsize}{!}{\includegraphics[]{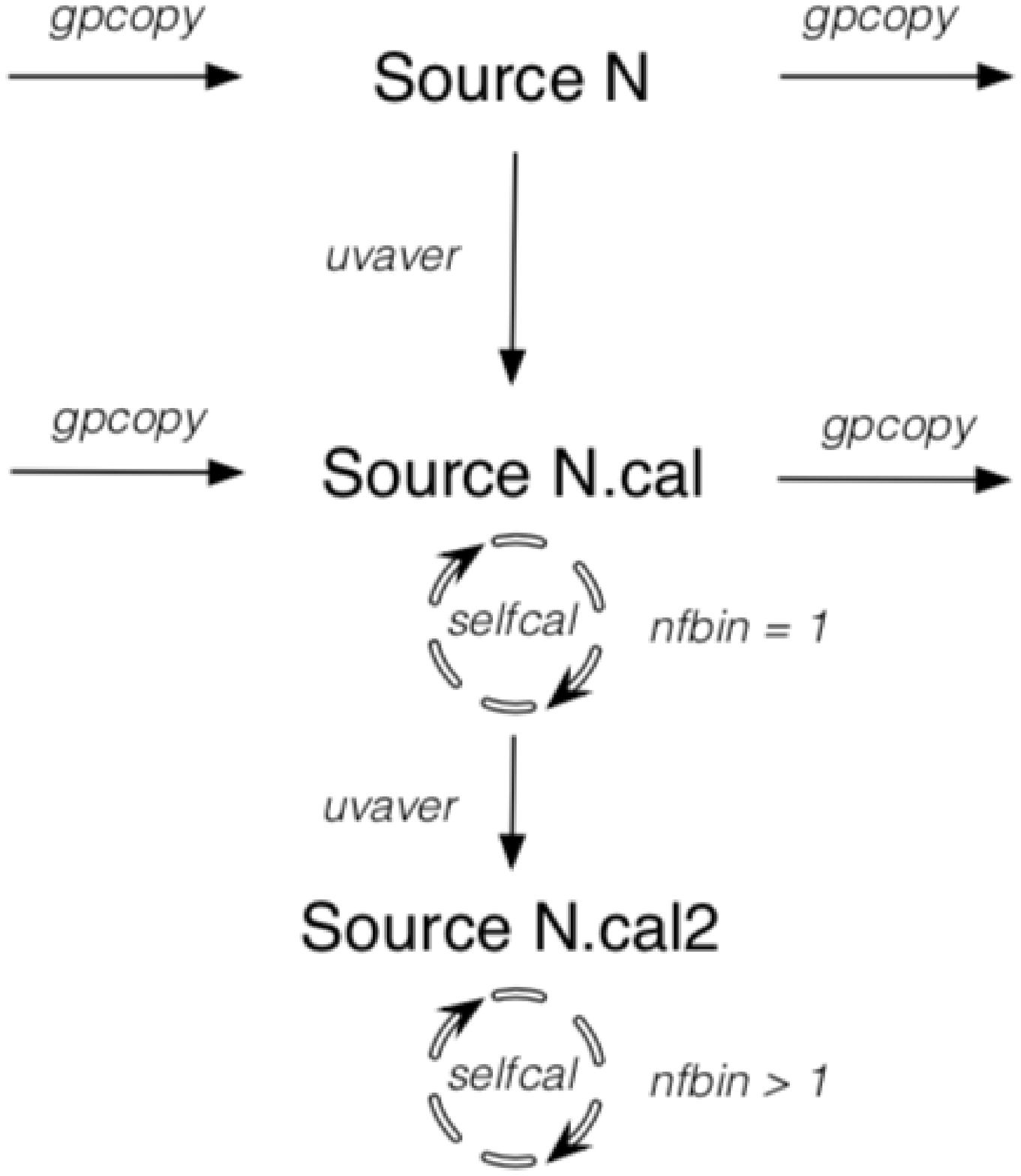}}
\caption{
Illustration of the self-calibration process for each target field, and how calibration solutions are transferred between fields using the $\textsc{miriad}$ task $\textsc{gpcopy}$. 
The task $\textsc{uvaver}$ applies calibration tables to the data, creating a new file in the process.
In the selfcal process, `nfbin' indicates into how many subbands the frequency band is split. 
Not shown is how the calibration tables from calibrator sources are copied and applied to Source~N.cal.
For calibrators the lower pair of $\textsc{gpcopy}$s is missing.
}
\label{calstrategy.fig}
\end{centering}
\end{figure}

After all calibrators from a single observing run have been self-calibrated, we combine the calibration solutions into two big tables: one table contains solutions when the entire band is used in $\textsc{selfcal}$, the other table contains the solutions when we split the band into four subbands.
These two calibration tables have the same structure as the single table that is created during a standard observing run with the ATCA, when one calibrator is observed repeatedly.
The subsequent calibration of a target field is illustrated in Fig.~(\ref{calstrategy.fig}).
First, we transfer the calibration solution from PKS B1934-638 to this field; these solutions are stored in files `Source N', `Source N-1', etc., without being modified.
Then we apply the calibration solutions from PKS B1934-638 to the data using the task $\textsc{uvaver}$, creating a new file `Source N.cal'.
In the next two steps we copy the two big calibration tables (which contain gain amplitudes and phase corrections) from the calibrators, and we run $\textsc{uvaver}$ after we copied each of these tables.
This updates the file `Source N.cal' with the selfcal solutions from the calibrators.
Once this is done, we copy and apply the calibration solution from the last field that was self-calibrated successfully; often this is the field `Source N-1.cal' which was observed only moments before.
Before starting the self-calibration process, we check that `Source N.cal' contains at least one source that is brighter than 200~mJy in Stokes~$I$.
If the field contains only fainter sources, then we rely on the calibration tables from the calibrators and from the source that was self-calibrated most recently to calibrate the current target field.
If the self-calibration process is completed successfully, the calibration tables that we just calculated will help calibrate the field that is observed next, as Fig.~(\ref{calstrategy.fig}) shows.
After the self-calibration process finishes, we do not test if the source is unresolved on the sky. Given the large size of the synthesized beam ($\sim 2\arcmin\times1\arcmin$, Fig.~\ref{beamsizes.fig}), this will be the case for most sources.
The equivalent of Fig.~(\ref{calstrategy.fig}) for self-calibrating calibrator sources is only slightly different: we did not transfer selfcal solutions between calibrators, therefore, for calibrator sources the lower set of $\textsc{gpcopy}$s is missing.

\section{Data for targets observed on multiple days}\label{Appendix_B.sec}
\begin{table*}
\centering
\caption{
Overview of fitted Stokes $I$ flux densities (in mJy) at 2.1 GHz for sources that were observed on multiple days. 
`--' indicates that the source was not observed that day, or that the fit to the data was not reliable. Runs 1-4 take place on consecutive days, run 5 after (almost) 4 months.
}
\label{mult_days_Iref.tab}
\begin{tabular}{l*5{r@{ $\pm$ }l}}
\hline
Source 	& \multicolumn{2}{c}{run 1}		&	\multicolumn{2}{c}{2}	&	\multicolumn{2}{c}{3}	&	\multicolumn{2}{c}{4}	&	\multicolumn{2}{c}{5} \\
\hline
PKS B0420-014 &    3058 &       0 &    3047 &       0 &    3002 &       0 &    3075 &       0 & \multicolumn{2}{c}{--} \\
PKS B0537-441 & \multicolumn{2}{c}{--} &    7519 &       0 & \multicolumn{2}{c}{--} &    7447 &       0 & \multicolumn{2}{c}{--} \\
PKS B0607-157 & \multicolumn{2}{c}{--} & \multicolumn{2}{c}{--} &    3023 &       0 &    3104 &       0 & \multicolumn{2}{c}{--} \\
PKS B0823-500 &    6093 &       0 &    5882 &       2 &    6019 &       0 &    6158 &       0 &    5665 &       0 \\
PKS B1127-145 &    4304 &       0 &    4298 &       0 &    4242 &       0 &    4308 &       1 & \multicolumn{2}{c}{--} \\
PKS B1215-457 &    3854 &       0 &    3843 &       0 & \multicolumn{2}{c}{--} & \multicolumn{2}{c}{--} & \multicolumn{2}{c}{--} \\
PKS B1308-220$^1$ &    3343 &       0 &    3327 &       2 &    3272 &       0 &    3371 &       0 & \multicolumn{2}{c}{--} \\
PKS B1421-490 &    7378 &       2 & \multicolumn{2}{c}{--} & \multicolumn{2}{c}{--} &    7305 &       0 & \multicolumn{2}{c}{--} \\
PKS B1613-586 &    4626 &       0 &    4584 &       1 &    4630 &       0 &    4612 &       0 & \multicolumn{2}{c}{--} \\
PKS B1827-360$^2$ &    4294 &       0 &    4256 &       0 &    4268 &       0 & \multicolumn{2}{c}{--} & \multicolumn{2}{c}{--} \\
PKS B1934-638$^2$ &   12343 &       0 &   12329 &       0 &   12400 &       0 & \multicolumn{2}{c}{--} &   12261 &       0 \\
PKS B2032-350 & \multicolumn{2}{c}{--} & \multicolumn{2}{c}{--} &    3960 &       1 &    3967 &       0 &    4043 &       1 \\
\hline
\end{tabular}
\begin{list}{}{}
\item[$^1$] PKS B1308-220 was detected with a polarization percentage $>$ 0.1 per cent only on run 2, the signal measured on all other runs could therefore be purely instrumental.
\item[$^2$] the average polarization percentage of this source is $<$ 0.1 per cent, the measured polarized signal could therefore be purely instrumental.
\end{list}
\end{table*}
\begin{table*}
\centering
\caption{
Same as Table~\ref{mult_days_Iref.tab}, showing instead the intrinsic polarized flux density at 2.1 GHz (in mJy) before any depolarization takes place. If the model that describes the polarization measurements best consists of multiple components, then we list the properties of the brightest component.
}
\label{mult_days_L1ref.tab}
\begin{tabular}{l*5{r@{ $\pm$ }l}}
\hline
Source 	& \multicolumn{2}{c}{run 1}		&	\multicolumn{2}{c}{2}	&	\multicolumn{2}{c}{3}	&	\multicolumn{2}{c}{4}	&	\multicolumn{2}{c}{5} \\
\hline
PKS B0420-014 &    59.2 &     0.1 &    46.7 &     0.1 &    57.0 &     0.1 &    65.9 &     0.1 & \multicolumn{2}{c}{--} \\
PKS B0537-441 & \multicolumn{2}{c}{--} &   249.5 &     0.1 & \multicolumn{2}{c}{--} &   215.6 &     0.1 & \multicolumn{2}{c}{--} \\
PKS B0607-157 & \multicolumn{2}{c}{--} & \multicolumn{2}{c}{--} &    97.2 &     0.1 &    95.5 &     0.1 & \multicolumn{2}{c}{--} \\
PKS B0823-500 &    19.9 &     0.1 &    19.6 &     0.2 &    19.9 &     0.1 &    20.5 &     0.1 &    18.2 &     0.1 \\
PKS B1127-145 &   159.7 &     0.1 &   158.6 &     0.1 &   161.1 &     0.1 &   157.5 &     0.1 & \multicolumn{2}{c}{--} \\
PKS B1215-457 &    15.8 &     0.1 &    17.9 &     0.1 & \multicolumn{2}{c}{--} & \multicolumn{2}{c}{--} & \multicolumn{2}{c}{--} \\
PKS B1308-220$^1$ &     3.1 &     0.1 &   111.2 &     1.2 &     2.2 &     0.1 &     2.2 &     0.1 & \multicolumn{2}{c}{--} \\
PKS B1421-490 &   119.7 &     0.2 & \multicolumn{2}{c}{--} & \multicolumn{2}{c}{--} &    71.9 &     0.1 & \multicolumn{2}{c}{--} \\
PKS B1613-586 &    50.1 &     0.1 &    49.1 &     0.1 &    51.0 &     0.1 &    50.8 &     0.1 & \multicolumn{2}{c}{--} \\
PKS B1827-360$^2$ &     2.1 &     0.1 &     3.2 &     0.1 &     3.0 &     0.1 & \multicolumn{2}{c}{--} & \multicolumn{2}{c}{--} \\
PKS B1934-638$^2$ &     4.4 &     0.1 &     6.2 &     0.1 &     5.5 &     0.1 & \multicolumn{2}{c}{--} &     2.6 &     0.1 \\
PKS B2032-350 & \multicolumn{2}{c}{--} & \multicolumn{2}{c}{--} &   242.7 &     0.2 &   264.7 &     0.1 &   260.8 &     0.2 \\
\hline
\end{tabular}
\end{table*}
\begin{table*}
\centering
\caption{
Same as Table~\ref{mult_days_L1ref.tab}, but for the RM (in \radm) of the brightest source component that was fitted to the polarization data. 
}
\label{mult_days_RM1.tab}
\begin{tabular}{l*5{r@{ $\pm$ }l}}
\hline
Source 	& \multicolumn{2}{c}{run 1}		&	\multicolumn{2}{c}{2}	&	\multicolumn{2}{c}{3}	&	\multicolumn{2}{c}{4}	&	\multicolumn{2}{c}{5} \\
\hline
PKS B0420-014 &   -40.5 &     0.1 &   -46.7 &     0.1 &   -42.1 &     0.3 &   -47.7 &     1.2 & \multicolumn{2}{c}{--} \\
PKS B0537-441 & \multicolumn{2}{c}{--} &    59.7 &     0.0 & \multicolumn{2}{c}{--} &    60.6 &     0.0 & \multicolumn{2}{c}{--} \\
PKS B0607-157 & \multicolumn{2}{c}{--} & \multicolumn{2}{c}{--} &    68.0 &     0.2 &    70.0 &     0.2 & \multicolumn{2}{c}{--} \\
PKS B0823-500 &   357.8 &     0.3 &   350.8 &     0.5 &   362.9 &     0.3 &   361.8 &     0.3 &    -2.0 &     0.2 \\
PKS B1127-145 &    42.6 &     0.1 &    41.1 &     0.1 &    41.4 &     0.1 &    43.7 &     0.1 & \multicolumn{2}{c}{--} \\
PKS B1215-457 &  -121.3 &     0.8 &  -114.7 &     0.5 & \multicolumn{2}{c}{--} & \multicolumn{2}{c}{--} & \multicolumn{2}{c}{--} \\
PKS B1308-220$^1$ &   -65.1 &     1.8 &     0.3 &     0.2 &    97.0 &     2.2 &   180.5 &     3.0 & \multicolumn{2}{c}{--} \\
PKS B1421-490 &    20.5 &     1.5 & \multicolumn{2}{c}{--} & \multicolumn{2}{c}{--} &    33.6 &     0.5 & \multicolumn{2}{c}{--} \\
PKS B1613-586 &    63.7 &     0.2 &    57.0 &     0.2 &    62.0 &     0.2 &    59.9 &     0.2 & \multicolumn{2}{c}{--} \\
PKS B1827-360$^2$ &  -585.6 &     1.9 &  -222.9 &     1.9 &  -206.8 &     1.6 & \multicolumn{2}{c}{--} & \multicolumn{2}{c}{--} \\
PKS B1934-638$^2$ &   575.7 &     1.7 &   498.6 &     1.0 &   594.3 &     1.5 & \multicolumn{2}{c}{--} & -2404.0 &     2.8 \\
PKS B2032-350 & \multicolumn{2}{c}{--} & \multicolumn{2}{c}{--} &     1.6 &     0.0 &     4.7 &     0.1 &     2.0 &     0.0 \\
\hline
\end{tabular}
\end{table*}

\bsp	\label{lastpage}
\end{document}